\documentclass[aps,prb,nobibnotes,twocolumn,superscriptaddress]{revtex4-1}
\usepackage{graphicx}
\usepackage{bm}
\usepackage{amsmath}
\usepackage{amssymb}
\usepackage{euscript}
\usepackage{verbatim}
\usepackage{fancyhdr}
\usepackage{wrapfig}
\usepackage{setspace}
\usepackage{xcolor}
\usepackage{amsfonts}
\usepackage{subfigure}
\usepackage{array}

\usepackage{hyperref}

\newcommand{\Renyi}[0]{R\'{e}nyi}
\newcommand{\Tr}[1]{\mathrm{Tr} #1}

\newcommand{\ket}[1]{\left| #1 \right\rangle}
\newcommand{\bra}[1]{\left\langle #1 \right|}
\newcommand{\tket}[1]{| #1 \rangle}
\newcommand{\tbra}[1]{\langle #1 |}
\newcommand{\Trt}[0]{\widetilde{\mathrm{Tr}}}

\begin{document}


\title{Entanglement spectroscopy of non-Abelian anyons: Reading off quantum dimensions of individual anyons}

\author{Eyal Cornfeld}
\affiliation{Raymond and Beverly Sackler School of Physics and Astronomy, Tel-Aviv University, Tel Aviv 6997801, Israel}
\affiliation{Department of Condensed Matter Physics, Weizmann Institute of Science, Rehovot 7610001, Israel}

\author{L. Aviad Landau}
\affiliation{Raymond and Beverly Sackler School of Physics and Astronomy, Tel-Aviv University, Tel Aviv 6997801, Israel}

\author{Kirill Shtengel}
\affiliation{Department of Physics and Astronomy, University of California, Riverside CA 92511, USA}

\author{Eran Sela}
\affiliation{Raymond and Beverly Sackler School of Physics and Astronomy, Tel-Aviv University, Tel Aviv 6997801, Israel}
\affiliation{Department of Physics and Astronomy and Quantum Materials Institute, University of British Columbia, Vancouver, B.C., V6T 1Z1, Canada }

\begin{abstract}
We study the entanglement spectrum of topological systems hosting non-Abelian anyons. Akin to energy levels of a Hamiltonian, the entanglement spectrum is composed of symmetry multiplets. We find that the ratio between different eigenvalues within one multiplet is universal and is determined by the anyonic quantum dimensions. This result is a consequence of the conservation of the total topological charge. For systems with non-Abelian topological order, this generalizes known degeneracies of the entanglement spectrum, which are hallmarks of topological states. Experimental detection of these entanglement spectrum signatures may become possible in Majorana wires using multicopy schemes, allowing the measurement of quantum entanglement and its symmetry resolution.
\end{abstract}

\maketitle
	


\section{Introduction and main result}\label{sec:intro}
Quantum entanglement gained a pivotal role in our understanding of many-body systems and 
specifically topological order~\cite{PhysRevLett.96.110404,PhysRevLett.96.110405,jiang2012identifying}.~
Its simplest measure is the entropy of entanglement between region \(A\) and region \(B\) of a bipartitioned system. In terms of the reduced density matrix of a subsystem $A$, $\rho_A = \rm{tr}_B \rho$, the entanglement entropy  is given by the  associated entropy $\mathcal{S} =  - \rm{Tr} (\rho_A \log \rho_A)$. Much more information however is encoded in the entanglement spectrum (ES), consisting of the spectrum $\{ \lambda \}$ of eigenvalues of $\rho_A$. In topological phases, the ES provides valuable information on the physics of the  edge. One may relate the ES to energy levels of an ``entanglement Hamiltonian"~\cite{Li2008} living on the edge of region $A$, defined via $\rho_A = e^{-H_E}$, or $\lambda_i =  e^{- \varepsilon_i}$. 

The significance of the ES in topological phases was introduced in the seminal paper of Li and Haldane in 2008 which studied 2D fractional quantum Hall (FQH) ground states~\cite{Li2008}. The low lying part of the ES was found to exhibit universal features related with the conformal field theory spectra of edge states living on the entanglement cut. This bulk-boundary correspondence was then extensively explored in a large body of studies~\cite{chandran2011bulk,qi2012general,poilblanc2012topological,dubail2012real,sterdyniak2012real,estienne2013matrix,yan2019bulk} testing a variety of models such as fractional Chern insulators~\cite{regnault2011fractional}, resonating valence bond wave functions~\cite{poilblanc2012topological}, or topological spin liquids~\cite{poilblanc2015chiral}; for a recent review see Ref.~\onlinecite{laflorencie2016quantum}. Along with these findings it became clear that the ES is a natural quantity for powerful  numerical algorithms, based on, \textit{e.g.}, matrix product states or projected entangled pair states (PEPS). This general understanding of the role of the ES in topological systems included both Abelian states (\textit{e.g.} the $\nu=1/3$ FQH state) and non-Abelian states (\textit{e.g.} the $\nu=5/2$ Moore-Read state or a $p+ip$ superconductor). Here, we take a closer look at the ES in non-Abelian states. We make progress using a topological quantum field theory language, specifically building on a 1D perspective.

The perspective of entanglement developed into a central approach towards  classification of interacting topological phases, with powerful results specifically in 1D systems~\cite{fidkowski2010effects,pollmann2010entanglement,turner2011topological,fidkowski2011topological,chen2011classification}, including symmetry protected topological (SPT) states and generalizations to higher dimensions~\cite{chen2013symmetry,PhysRevB.84.165139}. 
Direct manifestations of topological phases have been identified in 1D as \emph{symmetry-protected degeneracies} in the ES~\cite{turner2011topological} which are fingerprints of topological states in many different models~\cite{fidkowski2010entanglement,poilblanc2010entanglement,huang2011topological,deng2011entanglement,li2013identifying,duivenvoorden2013topological,nonne2013symmetry,morimoto2014z}.
These degeneracies are a consequence of the action of symmetry on the eigenvectors of the reduced density matrix $\rho_A$ (the ``Schmidt states").
A beautiful argument~\cite{turner2011topological} asserts that in gapped 1D systems with a finite correlation length $\xi$, every symmetry operation acts as a unitary operator on the ``low energy" Schmidt states, and can be factorized into two operators $\mathcal{O}_L \mathcal{O}_R$, acting \emph{locally} within distance $\xi$ from the left or right ends of segment $A$, respectively. Thus the eigenstates of $H_E$ provide a representation of the symmetry group. 
To illustrate this  Ref.~[\onlinecite{turner2011topological}] considered the Haldane integer-spin chain. Any $SU(2)$ spin rotation can be represented on the low lying eigenstates of $H_E$ in terms of generators $S_L$ and $S_R$ acting near the two ends of region $A$, where $S_{tot}=S_L+S_R$.
Since $S_{tot}$ is integer, we can have two possibilities: a ``trivial" phase in which both $S_L$ and $S_R$ are integer, or a ``topological" phase with both $S_L$ and $S_R$ half integer. In the latter case there must be an ES degeneracy of at least 4, due to a the two-fold degeneracy associated with a half-integer spin \emph{at each end}.

This 
should be contrasted with systems hosting non-Abelian anyons on their edges, like  the Kitaev chain~\cite{Kitaev2001}  with $\mathbb{Z}_2$ parity-number conservation, or parafermion chains~\cite{alicea2016topological}, where the edge operators do not support a local Hilbert space. For example in the Kitaev chain the symmetry operator measuring the total parity decomposes as $\mathcal{P}_{tot} = i \gamma_L \gamma_R$, where $\gamma_{L/R}$ are Majorana fermion modes, composed of degrees of freedom near either left or right end, which however
do not support a local Hilbert space. This is a common feature of non-Abelian anyons.

Generally, in the presence of symmetry, the reduced density matrix and hence $H_E$ can be block-diagonalized according to symmetry quantum numbers~\cite{laflorencie2014spin,goldstein2017symmetry,PhysRevB.98.041106}. Such a charge-resolution was recently applied in two-dimensional topological phases~\cite{matsuura2016charged} allowing to generalize the topological entanglement entropy~\cite{PhysRevLett.96.110404,PhysRevLett.96.110405} for systems protected by a symmetry.

In this paper we rely on this general symmetry-decomposition scheme, applied for the case of topological charge, defined below, in order to uncover internal structure in the ES in topological gapped phases supporting non-Abelian anyons. We demonstrate our predictions in explicit 1D lattice models of interacting anyons, and show that one can test them in experiment within available experimental setups in the context 
of Majorana wires.

\subsection{Non-Abelian anyons}
The idea to use the multi-dimensional, non-local space spanned by non-Abelian
anyons to encode quantum information was put forward by Kitaev~\cite{kitaev2003fault}. 
Non-Abelian anyons are physically realized as quasiparticles in systems with topological order. Operations such as braiding and fusion form the basis for topological quantum computation; for a review see Ref.~[\onlinecite{Nayak2008Non-Abelian}]. The universal statistical properties of anyons are encoded in their topological quantum field theory (TQFT)~\cite{KITAEV20062,BONDERSON2017399,pachos2012introduction}.
Our subsequent symmetry decomposition of entanglement relies on the concept of ``topological charge" or ``total fusion channel", a conserved quantity in anyonic systems. Consider a collection of anyons taken from a set $\{ I,a,b,c \dots \}$ satisfying certain fusion rules $a \times b=\sum_c N_{ab}^c c$, meaning that two anyons $a$ and $b$ can ``fuse into" an anyon of type $c$ as long as the (integer valued) fusion coefficient $N_{ab}^c$ is finite. If $N_{aa}^c \ne 0$ for more than one fusion outcome $c$, then a collection of $a$ anyons can encode information non-locally, and anyon $a$ is said to be non-Abelian. The total topological charge of the collection of anyons can be obtained by sequentially fusing pairs of anyons ending with a single anyon $a_{tot}$. For $N$ anyons of type $a$, the number of intermediate fusion channels grows exponentially as $d_a^{N}$, where $d_a$ is the \emph{quantum dimension}.
Different choices of a sequence of fusion are related by a basis transformation. On the other hand the total fusion outcome is a conserved ``topological charge".

\begin{figure} 
\centering	
\includegraphics[width=1\linewidth]{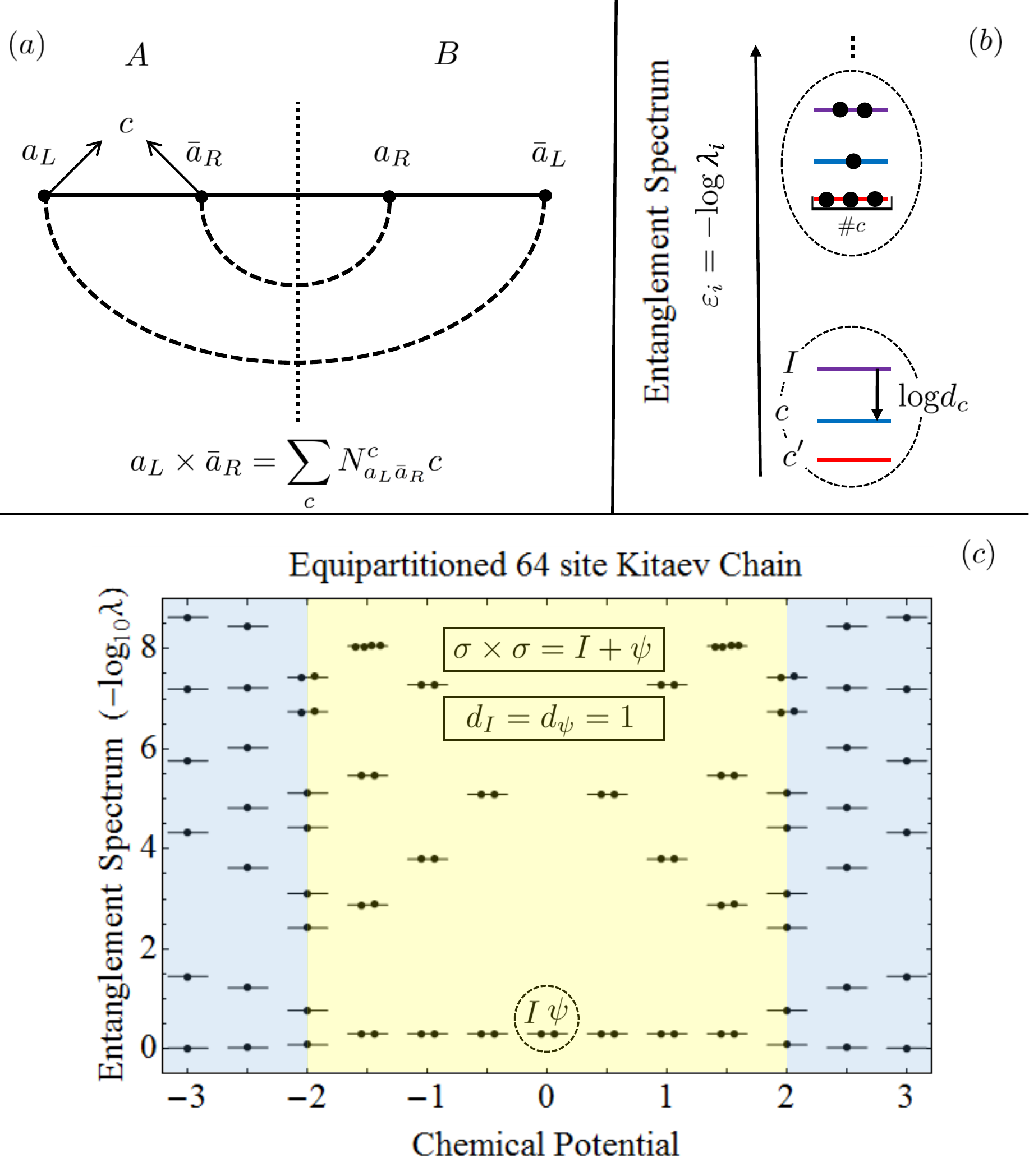}
\caption{(a) Bipartition of a 1D system on an open segment. The wave function admits a Schmidt decomposition in terms of the boundary anyon pair $a_L,\bar{a}_L$ and a pair $a_R,\bar{a}_R$  at the cut between $A$ and $B$. The total charge in region $A$, $c=a_A$, is determined by the fusion of the anyons in $A$. (b) Universal structure of the ES composed of many multiplets described by Eq.~(\ref{structure}) with internal multiplicities and spacing determined by the quantum dimension of $c$. (c) ES of the Kitaev chain $H = \sum_{j=1}^{64-1} (c^\dagger_j c_{j+1} + c^\dagger_j c^\dagger_{j+1} + h.c. )+ \mu\sum_{j=1}^{64} c^\dagger_j c_{j}$ for chemical potential $\mu = -3, -2.5, -2, \dots, 3$. Here $c= 1 , \psi$ takes two values with the same quantum dimension. The resulting topological degeneracy is removed at the transition, $|\mu| \ge 2$. At
the sweet spot 
$\mu=0$ there is a single multiplet.
} 
\label{fig:1}
\end{figure}

To study the entanglement between subsystems $A$ and $B$ in some topological ground state $|  \Psi \rangle$,  
it is natural to work in a basis where one fuses all anyons in each region into their total fusion channel $a_{A}$ and $a_{B}$ respectively. Then the total topological charge $a_{tot}$ is determined by $a_{A} \times a_{B} =\sum_{a_{tot}} N_{a_{A} a_{B}}^{a_{tot}} a_{tot}$. While the full wave function $|  \Psi \rangle$ has a well defined topological charge, importantly entangled states are a superposition of different sub-system charges $a_{A} , a_{B}$.
We restrict our attention to the natural case of a closed system with a total topological charge corresponding to the vacuum state, \textit{i.e.}, $a_{tot}=I$. 
In what follows we reserve the letter $c$ to denote the total fusion channel in region $A$, $c=a_A$, and assume $a_B = \bar{c}$.
The conservation of the topological charge implies~\cite{singh2014matrix} that $\rho_A$ forms blocks labeled by $c$.
Hence the ES is a union of the ES of each block with fixed topological charge. One may ask: what is the contribution to the entanglement from each charge sector~\cite{goldstein2017symmetry}? is there a general relation between these contributions?

For concreteness we consider 1D gapped models on an open segment.
Suppose that our 1D topological phase
carries an anyon of type $a$
as edge states. 
One may then postulate that the wave function $|  \Psi \rangle$ is described 
in terms of pair of anyons of type $a$ pulled out of the vacuum across each entanglement cut between $A$ and $B$. 
For our 1D system on an open segment we have two such pairs of anyons, as depicted in Fig.~\ref{fig:1}(a). 
From this picture, using the rules of TQFT we find that the ES decomposes according to the subsystem topological charge $c$
as 
\begin{equation}
\label{structure}\{ \lambda\} = {\textstyle\bigcup\nolimits_{c}} \{ \lambda\}_{c},~~~\#_{c} = N_{aa}^{c}, ~~~\lambda_c \propto d_{c}.
\end{equation} 
The ES for each charge sector $c$ consists of a single eigenvalue $\lambda_{c}$ with  multiplicity $\#_{c}$.  The multiplicity is just the number of possibilities for fusing the boundary anyons into the total topological charge $c$, $\#_{c} = N_{aa}^{c}$. Our main result is that $\lambda_{c}$ probes the quantum dimension of the type-$a$ anyon, $\lambda_a \propto d_{a}$.
We can see that in general anyon systems the ES is characterized by universal \emph{ratios} of its eigenvalues determined by the quantum dimensions $d_{c}$, instead of degeneracies. Equivalently, multiplets in the entanglement Hamiltonian have the same decomposition $\{ \varepsilon\} = \cup_{c} \{ \varepsilon\}_{c}$  with multiplicities $\#_{c}$, and with an additive entropy term 
\begin{equation}
\label{deltaE}
\delta \varepsilon_{c}= - \log d_{c},
\end{equation}
see Fig.~\ref{fig:1}(b). This central result demonstrates that the charge-resolved entanglement contains TQFT data, which can be interpreted as the entropy associated with a single anyon $c$. 

A similar term appears~\cite{PhysRevLett.96.110404,PhysRevLett.96.110405} as the subleading term $\gamma$, referred to as topological entanglement entropy, in 2D topologically ordered systems with a finite correlation length $\mathcal{S}(L_A) = \alpha L_A  -  \gamma + ...$. Here $\alpha$ is a non-universal constant, while $\gamma = \log \mathcal{D} $  is a universal number known as the total quantum dimension $ \mathcal{D}= \sqrt{\sum_a d_a^2}$ characterizing the topological medium.

In generic models hosting non-Abelian anyons
there are additional local degrees of freedom creating entanglement on the scale of the correlation length $\xi$ across the entanglement cut. These are reflected in the ES by additional non-universal levels. When these local degrees of freedom are independent from the anyons, each such level becomes a multiplet with exactly the same structure as the ground state described by Eq.~(\ref{structure}), see dashed circle in Fig.~\ref{fig:1}(b).

The Kitaev chain realizes a familiar but unfortunately non-exhaustive example of Eq.~(\ref{structure}). It is described in terms of Ising anyons $\{ I,\psi,\sigma \}$, where for open boundary conditions the edges of the chain host $\sigma$-anyons (corresponding to $\gamma_L$ and $\gamma_R$ above),
with fusion rules $\sigma_L \times \sigma_R  = I  + \psi$. We thus have two topological charges  $I$ and $\psi$ corresponding to even and odd electron parity in region $A$, respectively. Accordingly the ES is decomposed as $\{\lambda \} = \{\lambda \}_I \cup \{\lambda \}_\psi  $.  
However,  both of anyons $I$ and $\psi$ are Abelian, $d_I = d_\psi = 1$ implying $\lambda_I = \lambda_\psi = 1/2$. This yields the familiar \emph{degeneracy} in the ES of the topological phase in the Kitaev model, seen in Fig.~\ref{fig:1}(c). Thus it would be interesting to explore the ES and the prediction in Eq.~(\ref{structure}) in models hosting more exotic anyons. 

The rest of the paper is organized as follows. After deriving the main result Eq.~(\ref{structure}) in Sec.~\ref{se:TQFT}, in Sec.~\ref{se:AKLT} we demonstrate its validity for specific lattice models. 
In contrast to the Kitaev model, in order to observe nontrivial ratios in the ES, one needs models whose edge states can fuse into a non-Abelian anyons having quantum dimension $d_a>1$. 
$\mathbb{Z}_n$ parafermions which have received considerable attention recently~\cite{alicea2016topological} are richer than Majorana fermion modes since their edge states have $n \ge 2$ fusion outcomes, which themselves, however, are all Abelian anyons.
Fibonnaci anyons satisfying the fusion rule $\tau \times \tau = 1+\tau$
are rich enough to display the nontrivial structure in the ES found here. Indeed, 1D models with exotic states have been built out of these anyons~\cite{feiguin2007interacting,trebst2008collective,trebst2008short}. Here, we demonstrate our results for $SU(2)_k$ anyonic generalizations of the AKLT chain~\cite{gils2013anyonic}.

In Sec.~\ref{se:exp} we discuss an experimental protocol to detect the topological charge resolution of the entanglement. While the ES can be measured in small systems, as was recently demonstrated for a small system realizing the AKLT chain~\cite{choo2018measurement}, here we develop methods for general systems. We focus on topological-charge resolution of entanglement measures. As a first step towards this goal, we focus on Majorana systems, specifically on their implementations in quantum wires controllable by charging energy effects~\cite{barends2014superconducting}. Various protocols have been proposed for designing entangled states~\cite{van2012coulomb,hassler2012strongly,aasen2016milestones,landau2016towards,litinski2018quantum} (please note that this is not a complete list).
In accordance with our results, the ES in particular and other entanglement measures in general can be decomposed into degenerate even/odd sectors. Using a general protocol for measurement of charge-resolved entanglement~\cite{goldstein2017symmetry,cornfeld2018imbalance,cornfeld2018measuring}, we discuss a concrete platform for measuring the second \Renyi{} entropy. In the topological phase hosting Majorana  edge states, the separately measurable even ($c=I$) and odd ($c=\psi$) contributions to the \Renyi{} entropy become degenerate. This allows to measure experimentally the predicted topological degeneracies in the ES and assign it to the charge resolution.

\section{Boundary-anyons Entanglement}
\label{se:TQFT}

In this section we derive Eq.~(\ref{structure}). Bonderson \textit{et~al}.~\cite{BONDERSON2017399} have recently applied TQFT methods to study entanglement of anyonic systems. Here we use these methods and focus on the ES. The impatient reader may skip this TQFT-based derivation directly to Eq.~\ref{main} at the end of this technical section, and then to the next sections where its significance is exemplified. 

\begin{figure*}[t]
	\centering
	\includegraphics[width=1\linewidth]{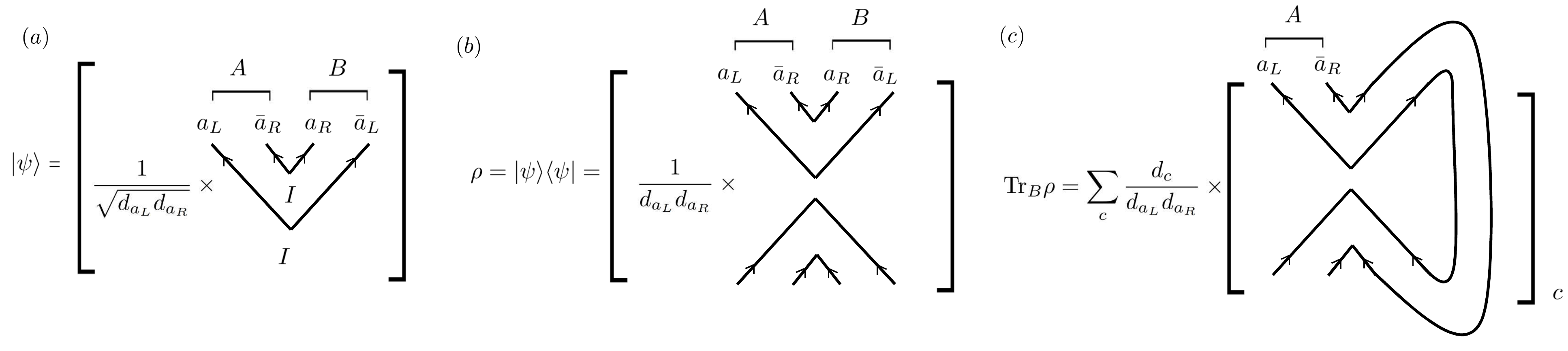}
	\caption{(a) Graphical representation of the normalized wave function (\textit{cf.} Ref.~[\onlinecite{BONDERSON2017399}]). Two pairs of anyons are fused out of the vacuum creating entanglement between $A$ and $B$. (b) Normalized density matrix. (c) Reduced density matrix,  separated into blocks with fixed total anyonic charge $c$ in region $A$. The diagram is equivalent to two parallel lines $[||]_c$ representing the identity operator in the Hilbert subspace of two anyons $a_L,\bar{a}_R$, fused into $c$; see Eq.~(\ref{eq:paralelc}).  }\label{fig:secII}
\end{figure*}

We consider the setting depicted in Fig.~\ref{fig:1}(a), where anyons $\{ a\}$ appear on the entanglement cut
between regions $A$ and $B$ as well as on the physical boundaries. 
We start with a system with boundary anyons $ a,\bar{a}$ pulled out of the vacuum at the entanglement cut between regions \(A\) and \(B\), whose joint state is represented by $|a, \bar{a};I \rangle^R$; a second pair of anyons at the real boundary of the system is represented by $|a, \bar{a};I \rangle^L$. The total system is in state $|\Psi \rangle = \tket{a,\bar{a};I}^L \tket{a,\bar{a};I}^R$ where $L,R$ correspond to the left and right edges of region $A$; this state is depicted in Fig.~\ref{fig:secII}(a) (here, we utilize the diagrammatic notations by Kitaev~\cite{KITAEV20062} as presented in Ref.~[\onlinecite{BONDERSON2017399}]; see Appendix~\ref{app:notation}). Consequently, the density matrix of the full system is
\begin{equation}
\rho= \tket{a,\bar{a};I}^L\tket{a,\bar{a};I}^R\tbra{a,\bar{a};I}^L\tbra{a,\bar{a};I}^R,
\end{equation}
see Fig.~\ref{fig:secII}(b).
We then partition the anyons into two subsystems \(A=a_L,\bar{a}_R,B= a_R , \bar{a}_L\), and proceed by tracing out subsystem $B$. This leads to a diagram  representing the identity in the Hilbert space spanned by two anyons $a_L,\bar{a}_R$. Our main step here is facilitated by the  decomposition of this space according to the total fusion channel $c$. This is diagrammatically represented in Fig.~\ref{fig:secII}(c), and results in
\begin{equation}
\label{eq:paralelc}
\rho_A=\Tr_B{\rho}=\sum_{c} \frac{d_c}{ d_{a_L}d_{\bar{a}_R}} \left[\mathbb{I}_{a_L, \bar{a}_R}\right]_c,
\end{equation}
where the square brackets describe the identity operator in the two-anyon Hilbert subspace with total fusion channel $c$~\cite{BONDERSON2017399}. As discussed in Appendix~\ref{app:general}, this is given by
\begin{gather}
[\mathbb{I}_{a_L, \bar{a}_R}]_c=\sum_{\mu=1}^{N_{a_L\bar{a}_R}^c}\tket{a_L\bar{a}_R;c,\mu}\tbra{ a_L \bar{a}_R ;c,\mu},
\end{gather}
where generally $a_L$ and $\bar{a}_R$ can fuse into $c$ in multiple ways labelled by $\mu$. This pair of equations yields the spectrum and degeneracies in Eq.~(\ref{structure}),
\begin{equation}
\label{main}
\lambda_c=\frac{d_c}{ d_{a_L}d_{\bar{a}_R}},\qquad
\#_c=N_{a_L, \bar{a}_R}^c.
\end{equation}
The normalization of the density matrix follows from the fusion algebra
\begin{equation}
\Tr\rho_A=\sum_c\frac{N_{ a_L \bar{a}_R }^c d_c}{  d_{a_L}d_{\bar{a}_R}}=1.
\end{equation}

Below we demonstrate the result Eq.~(\ref{main}), illustrated in Fig.~\ref{fig:1}(b), in concrete models of interacting anyons.

\section{Anyonic chains} 
\label{se:AKLT}
To test the topological field theory results, in this section we study a lattice model of interacting anyons. Before focusing on a specific model in Sec.~\ref{se:AKLT1}, we first discuss generalities about the construction of the ES in anyonic models.

Consider a system of anyons, some of which are located in region $A$, and the rest in $B$. The structure of a factorizable Hilbert space  $\mathcal{H} = \mathcal{H}_A \otimes \mathcal{H}_B$, which is usually taken for granted, is actually not present, since each individual anyon does not posses a local Hilbert space. 
However, for every value of the total fusion channel $a_A$ and $a_B$ in each region, there is a local Hilbert space $\mathcal{H}_A^{a_A}$ and $\mathcal{H}^{a_B}_B$, respectively. Thus, for a specific value of the total charge $a_{tot}$, the Hilbert space is a \emph{direct sum} of factorizable spaces, consistent with the fusion rules $\mathcal{H} = \bigoplus_{a_A,a_B, N_{{\scriptstyle a_A}{\scriptscriptstyle a_B}}^{\scriptstyle a_{tot}} =1} \mathcal{H}^{a_A}_A \otimes \mathcal{H}^{a_B}_B$ (here for simplicity \(N_{ab}^c=0,1\) for all \(a,b,c\)).

\begin{figure} 
	\centering	
	\includegraphics[width=1\linewidth]{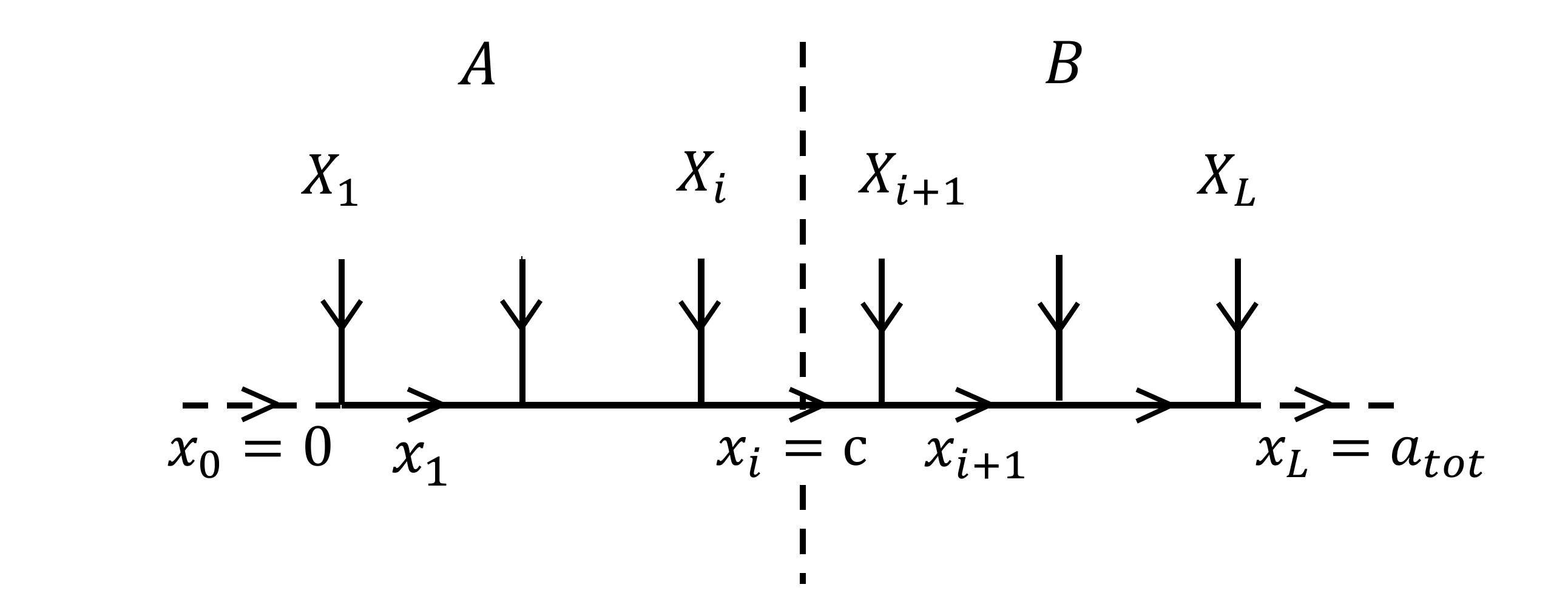}
	\caption{An open chain of $L$ anyons. We choose the total topological charge $a_{tot}=0$. The total topological charge of region $A$, $c = a_A$ is given by the link connecting regions $A$ and $B$.
	} 
	\label{fig:fusiondiagrammodel}
\end{figure}

Following Singh \textit{et~al}.~\cite{singh2014matrix} consider a chain of $L$ anyons denoted $\{X_i \}$ ($i=1,\dots,L$), whose state can be represented by the fusion diagram in Fig.~\ref{fig:fusiondiagrammodel}. A family of such 1D models was introduced in Refs.~[\onlinecite{feiguin2007interacting,trebst2008collective,trebst2008short,gils2013anyonic}]. This basis of states is denoted as \({\tket{x_1,\dots,x_{L-1}}^{X_1,\dots,X_{L}}_{x_0 x_L}=\tket{\{x\}}}^{\{X\}}_{x_0 x_L}\), with external legs \(\{X\}\) corresponding to the $L$ anyons, and with fixed boundary anyons $x_0$ and $x_L$. The fusion rule constraints \({N_{x_i,X_{i+1}}^{x_{i+1}}\neq 0}\) have to be satisfied at each vertex. Setting $x_0=I$ to the vacuum state, we can identify $x_L$ with the total fusion channel $a_{tot}$. We set it to the vacuum state $a_{L}=a_{tot}=I$. 

Consider dividing the chain into region $A$ containing anyons $X_1 \dots X_{i}$ and region  $B$ containing $X_{i+1} \dots X_{L}$. It is evident that the link $x_i$ connecting $A$ and $B$ carries the topological charge of region $A$,  hence we identify $x_i=a_A=c$. 
For each value of $c$, each basis vector may be partitioned into the two regions as \({\tket{\{a\},c,\{b\}}^{\{A\},\{B\}}=\tket{\{a\}}_{0c}^{\{A\}}\otimes\tket{\{b\}}_{c0}^{\{B\}}}\). This set of states with fixed boundary condition specified by $c$ defines the Hilbert space for $A$ and $B$, $\mathcal{H} = \bigoplus_{c} \mathcal{H}^{c}_A  \otimes \mathcal{H}^{\bar{c}}_B$ and allows us to write any state $\tket{\Psi}=\sum_{\{x\}}\Psi_{\{x\}}\tket{\{x\}}$ as 
\begin{align}
\tket{\Psi}=\sum_{c}\sum_{\{a\},\{b\}}[\Psi_{c}]_{\{a\},\{b\}}\tket{\{a\}}_{0c}^{\{A\}}\otimes\tket{\{b\}}_{c0}^{\{B\}}.
\end{align}
To obtain the ES one may treat the wavefunction $\Psi_{\{\{a\},c,\{b\}\}}$ as a matrix $[\Psi_c]_{\{a\},\{b\}}$ of (super-)indices \(\{a\},\{b\}\), representing the \(c\) component of the wavefunction, and perform a singular value decomposition (SVD) on that matrix,
\begin{align}
\label{eq:Psi1}
\tket{\Psi}=\sum_{c}\sum_{j}[\Lambda_{c}]_{j}\tket{\Phi^A_j}_{0c}^{\{A\}}\otimes\tket{\Phi^B_j}_{c0}^{\{B\}}.
\end{align}
Here, the vectors $\tket{\Phi^A_j}_{0c}^{\{A\}}$ and $\tket{\Phi^B_j}_{c0}^{\{B\}}$ labeled by $j$ belong to $\mathcal{H}^{c}_A$ and $\mathcal{H}^{\bar{c}}_B$, respectively. We now turn to an explicit example.

\subsection{Anyonic-AKLT chain} 
\label{se:AKLT1}
We consider the anyonic-$SU(2)_k$ AKLT chain~\cite{gils2013anyonic} as a concrete example.  The model 
contains a family of anyons labeled by their integer ``spin" quantum number $j=0,1,\dots ,  \lfloor k/2 \rfloor$. These are generally non-Abelian anyons, with fusion rules $j_1 \times j_2 = |j_1-j_2|, \dots, \min \{ j_1+j_2,k - j_1-j_2\}$, extending the usual $SU(2)$ rules. The $j=0$ anyon corresponds to the vacuum state $I$ with zero topological charge. The chain consists of $L$   anyons of type $j=1$, \textit{i.e.}, $X_i = 1$ ($i=1, \dots L$) and the $\{ x_i\}$ should be consistent with the fusion rules  $x_{i-1} \times X_{i}  = x_{i}$. The model contains an ``AKLT sweet spot" analogous to the familiar AKLT chain~\cite{affleck1987rigorous} with Hamiltonian $H_{\mathrm{sweet~spot}}=\sum_{i=1}^{L-1} P^{(2)}(X_i,X_{i+1})$, consisting of projectors of neighbouring $j=1$ anyons $X_{i},X_{i+1}$ onto the $j=2$ fusion channel. A parameter $\theta_{12}$ describes deviations from the AKLT sweet spot whereby an additional term $\propto \sum_i P^{(1)}(X_i,X_{i+1})$ appears in the Hamiltonian
\begin{equation}
\label{Htheta}
H\!=\!\sum\limits_{i=1}^{L-1} [P^{(2)}(X_i,X_{i+1})\cos\theta_{12} - P^{(1)}(X_i,X_{i+1})\sin\theta_{12}].
\end{equation} 
Using F-moves one can write the Hamiltonian in the basis of Fig.~\ref{fig:fusiondiagrammodel}~\cite{gils2013anyonic}; see Fig.~\ref{fig:Y}(b).

\begin{figure}[t]
	\includegraphics[width=1\linewidth]{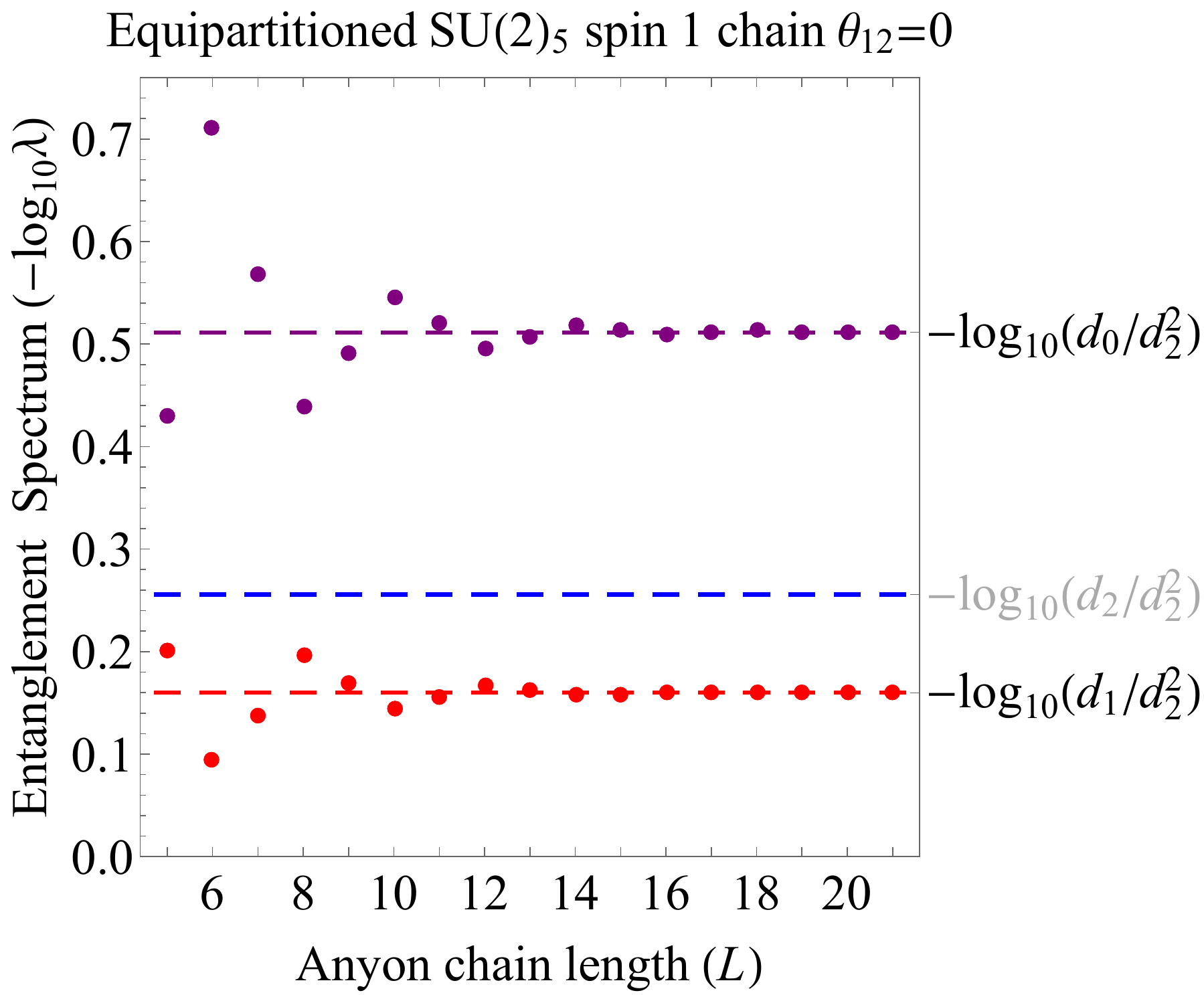}
	\caption{ES for the $SU(2)_5$ anyonic-AKLT ground states at $\theta_{12}=0$. Purple and red correspond to fusion channel $c=0,1$, respectively. Closed expression given at Eq.~(\ref{exactres}).}\label{ESsweetspot}
\end{figure}

\begin{figure}[t]
	\centering
	\includegraphics[width=1\linewidth]{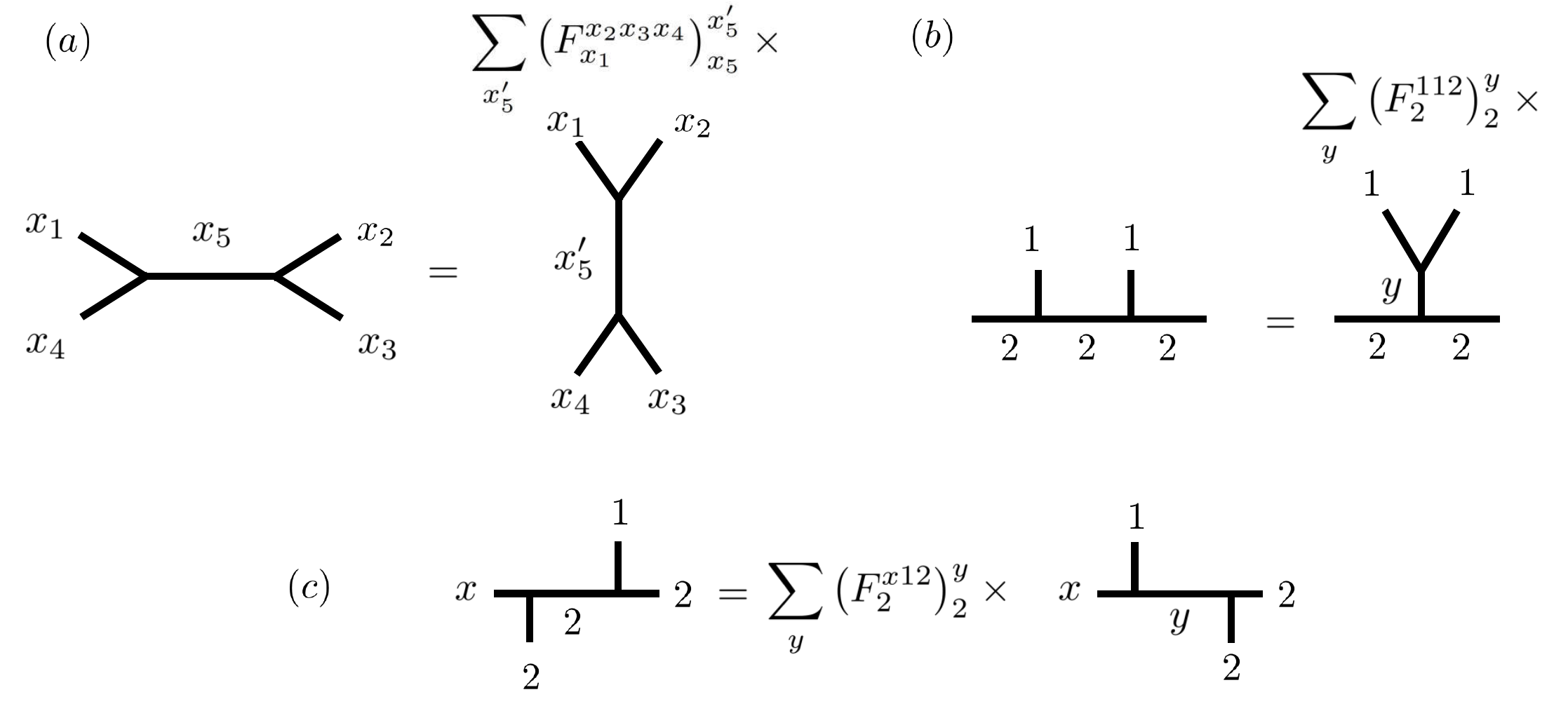}
	\caption{The F-move anyonic change of basis.}\label{fig:Y}
\end{figure}

We focus on the 
\(SU(2)_k\) model with odd 
$k \ge 5$~\cite{gils2013anyonic} and discuss explicitly the case of $k=5$, with 3 anyons types $j=0,1,2$ of quantum dimensions $d_j$
\begin{equation}
d_0=1,\qquad d_1=\frac{\sin(3\pi/7)}{\sin(\pi/7)},\qquad d_2=\frac{\sin(5\pi/7)}{\sin(\pi/7)}.
\end{equation}
These anyons satisfy the fusion rules
\begin{equation}
\label{eq:fusionrules}
1 \times 1 =0+1+2,\qquad  1 \times 2 =1+2,\qquad 
2 \times 2 = 0+1,
\end{equation}
along with $0 \times x =x$ ($x=0,1,2$).
The following results are obtained by exact diagonalization of this Hamiltonian, supplemented by analytic treatment at the AKLT sweet spot. 

\begin{figure*}[t]
	\centering
	\includegraphics[width=.6\linewidth]{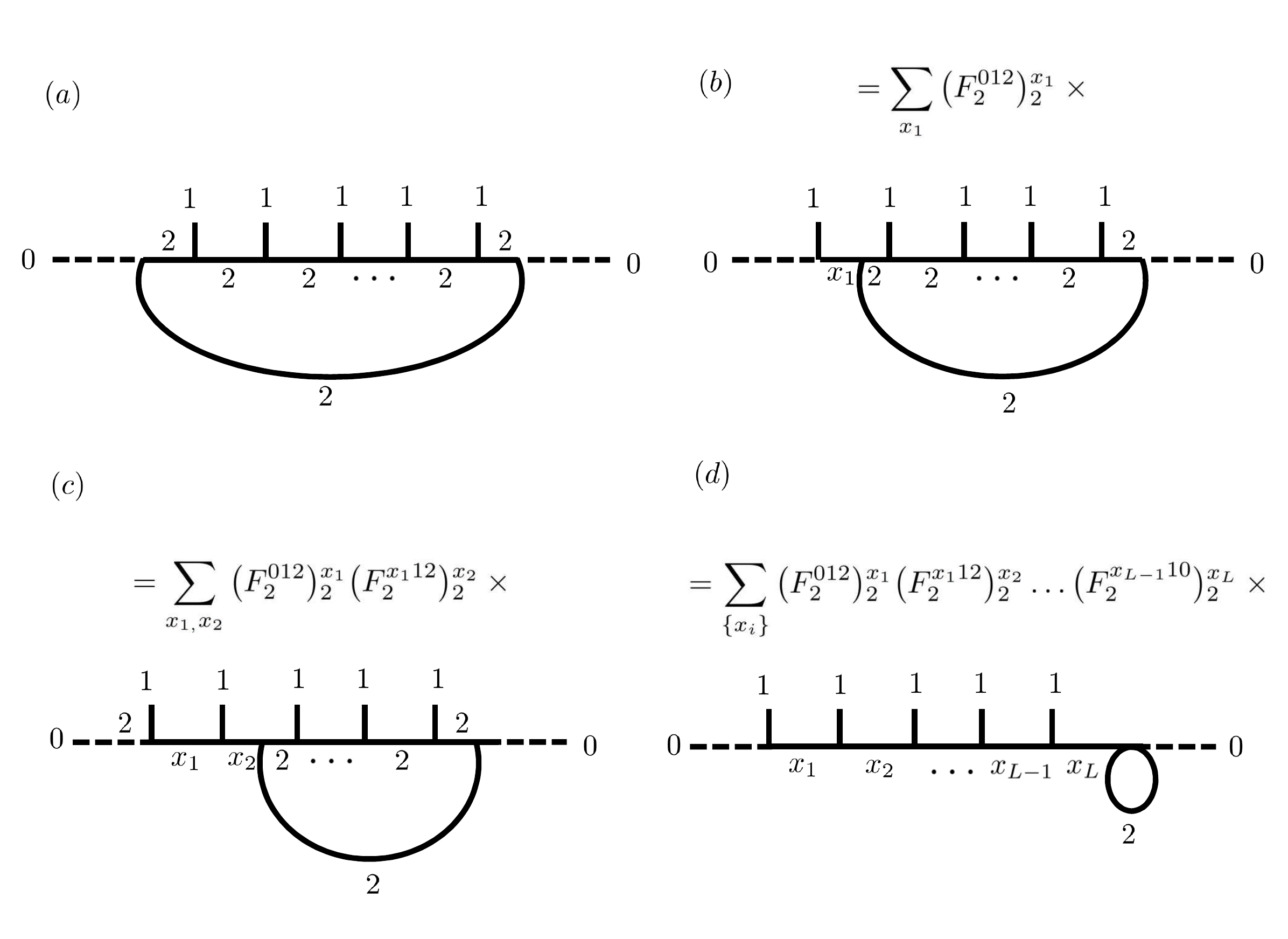}
	\caption{(a) Exact ground state for the AKLT anyonic chain with PBC~\cite{gils2013anyonic}. (b,c,d): Through a sequence of F-moves, the loop can be brought into contractible form. After its contraction one obtains the diagram with OBC Fig.~\ref{fig:fusiondiagrammodel}. This gives the exact ground state Eq.~(\ref{eq:psisqweet}). }\label{fig:X}
\end{figure*}

\subsubsection{Entanglement of the AKLT sweet spot}

The ES for an equipartition of the ground state of the open chain at the AKLT sweet spot is shown in Fig.~\ref{ESsweetspot} for various values of $L$. There are only two eigenvalues with a total fusion channel of region $A$ given by $c=0$ and $c=1$ (purple and red respectively). At sufficiently large $L$ their ratio approaches the ratio of the quantum dimensions $d_0/d_1$. This result may seem surprising in the absence of the underlying boundary anyons assumed in Fig.~\ref{fig:1}. Through an exact solution, however, we shall see that boundary anyons do indeed emerge and, accordingly, the resulting ES is in line with our main result, Eq.~(\ref{main}).

\subsubsection{Exact solution}
To find the exact ground state for the AKLT chain with open boundary conditions (OBC), we start from the exact ground state that can be readily guessed~\cite{gils2013anyonic} for periodic boundary conditions (PBC), see Fig.~\ref{fig:X}(a). The state depicted in Fig.~\ref{fig:X}(a) is a ground state since each pair of adjacent $j=2$ anyons can never fuse into $j=2$ according to the fusion rules Eq.~(\ref{eq:fusionrules}), yielding a zero energy state which is a ground state of the positive definite Hamiltonian $H_{\theta_{12}=0}$. Below we discuss its uniqueness. In order to obtain the corresponding ground state for OBC in the basis $|\{ x \} \rangle_{x_0=0,x_L=0}$ of Fig.~\ref{fig:fusiondiagrammodel}, we sequentially shrink and finally annihilate the loop. This is done by applying F-moves 
according to the rule in Fig.~\ref{fig:Y}(c), 
and is implemented in Fig.~\ref{fig:X}(b,c,d). This gives back the OBC diagram of Fig.~\ref{fig:fusiondiagrammodel}.
The resulting state is explicitly given in terms of the $F$-matrices~\cite{gils2013anyonic} by
\begin{equation}
\label{eq:psisqweet}
\Psi_{\{x_i \}} \propto \prod_{i=0}^{L-1} (F_2^{x_i ,1 ,2})_2^{x_{i+1}}.
\end{equation}
This is the unique ground state of the total topological sector $a_{tot}=0$. In comparison, the usual spin-1 AKLT Hamiltonian~\cite{affleck1987rigorous}
has 4 ground states, obtained by the singlet and triplets that the two boundary spin-1/2-s can form. In our anyonic model in Fig.~\ref{fig:fusiondiagrammodel} we have imposed a total anyonic charge 0, analogous to restricting ourselves to the the AKLT singlet state which is also unique.

As shown in Appendix~\ref{app:ESAKLT}, one can analytically obtain the ES from this state. Specifically, for a bipartition of a chain with \(L\) anyons to regions \(A\cup B\) consisting of \(L_A+L_B=L\) anyons, we find
\begin{eqnarray}\label{exactres}
\lambda_{0}&=&\frac{1}{d_{2}^2}\cdot\frac{(1-(-d_{1})^{1-L_A})(1-(-d_{1})^{1-L_B})}{(1-(-d_{1})^{1-L_A-L_B})}, \nonumber \\ 
\lambda_{1}&=&\frac{{d_{1}}}{d_{2}^2}\cdot\frac{(1-(-d_{1})^{-L_A})(1-(-d_{1})^{-L_B})}{(1-(-d_{1})^{1-L_A-L_B})}, \nonumber \\
\lambda_{2}&=&0.
\end{eqnarray}
One clearly sees that at the thermodynamic limit \(L\to\infty\) the spectrum corresponds with our main TQFT result, Eq.~(\ref{main}), and \(\lambda_{0,1}\to d_{0,1}^{\phantom{|}}/d_2^2\). This is plotted in Fig.~\ref{ESsweetspot} and matches the exact diagonalization numerics.

\subsubsection{Boundary anyons}
\begin{figure}[t]
	\centering
	\includegraphics[width=1\linewidth]{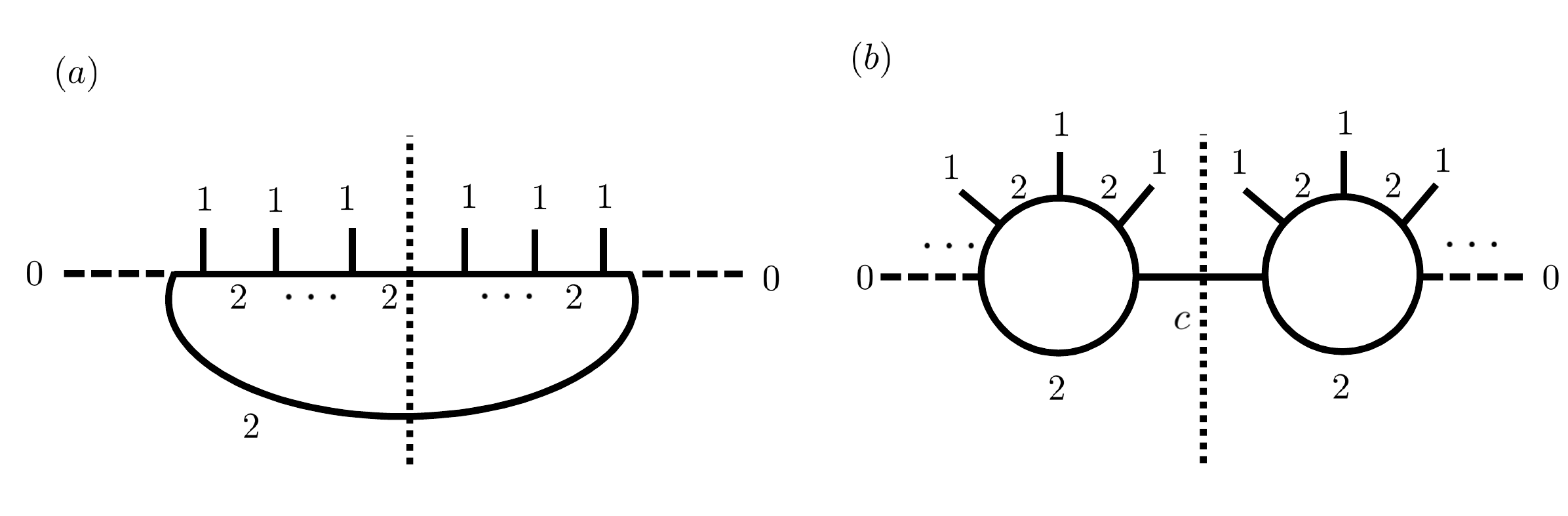}
	\caption{Visualisation of the ``hidden" boundary anyons of the OBC model depicted in Fig.~\ref{fig:fusiondiagrammodel} using the PBC representation of Fig.~\ref{fig:X}(a). }\label{fig:Z}
\end{figure}

In Fig.~\ref{fig:X}(a), which is the PBC representation of our OBC state, we can see two $j=2$ anyon lines connecting regions $A$ and $B$, see Fig.~\ref{fig:Z}. We associate them with the boundary anyons $a_L = a_R =\bar{a}_L =\bar{a}_R= 2$. Since the state is gapped, in the thermodynamic limit these boundary anyons are decoupled. 
The fusion rule $2 \times 2 = 0+1$ then dictates that the total charge of region $A$ can only take on values $c=1,2$, which explains the structure of the ES.

\begin{figure}[t]
	\includegraphics[width=1\linewidth]{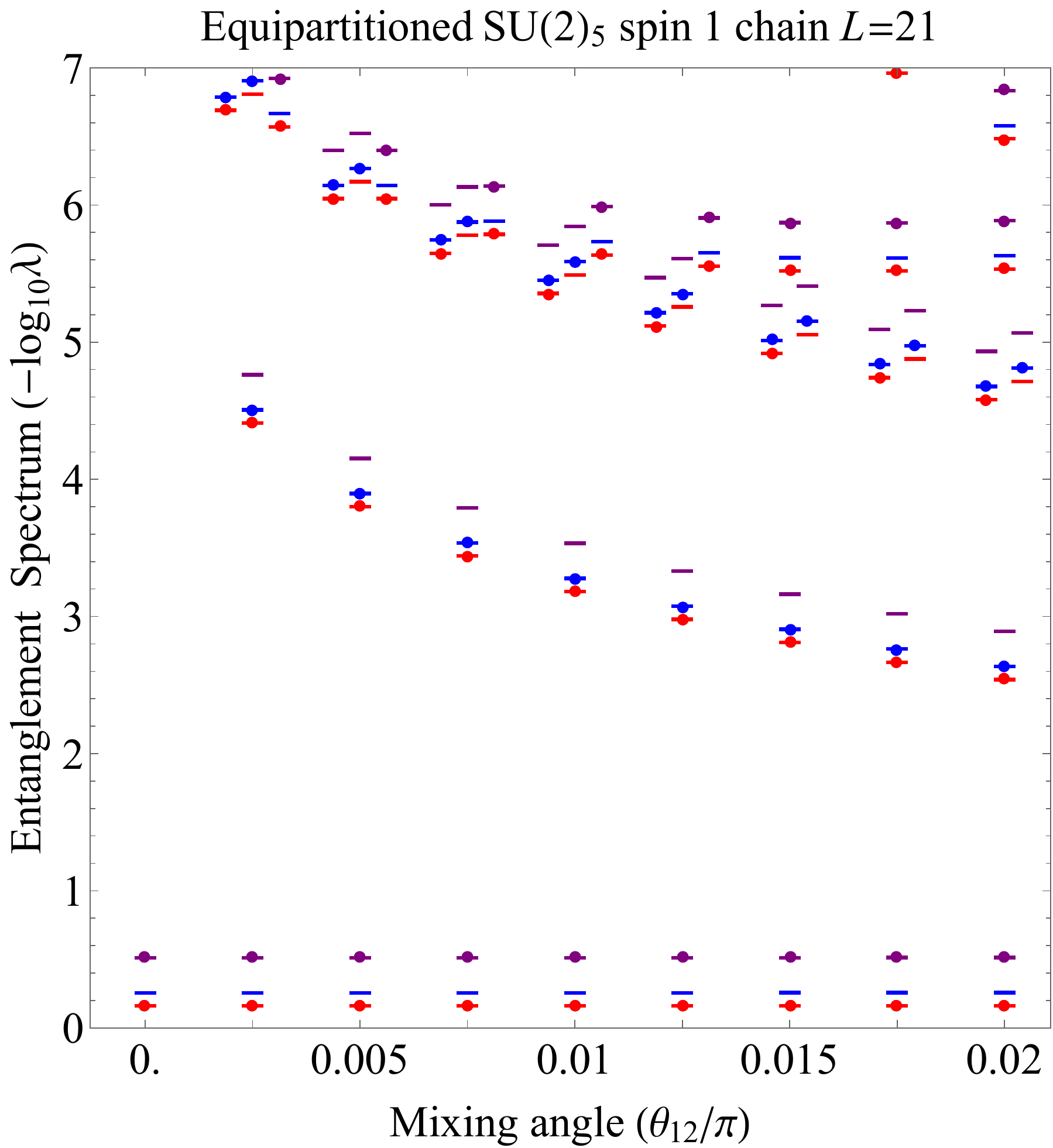}
	\bigskip\\
	\includegraphics[width=1\linewidth]{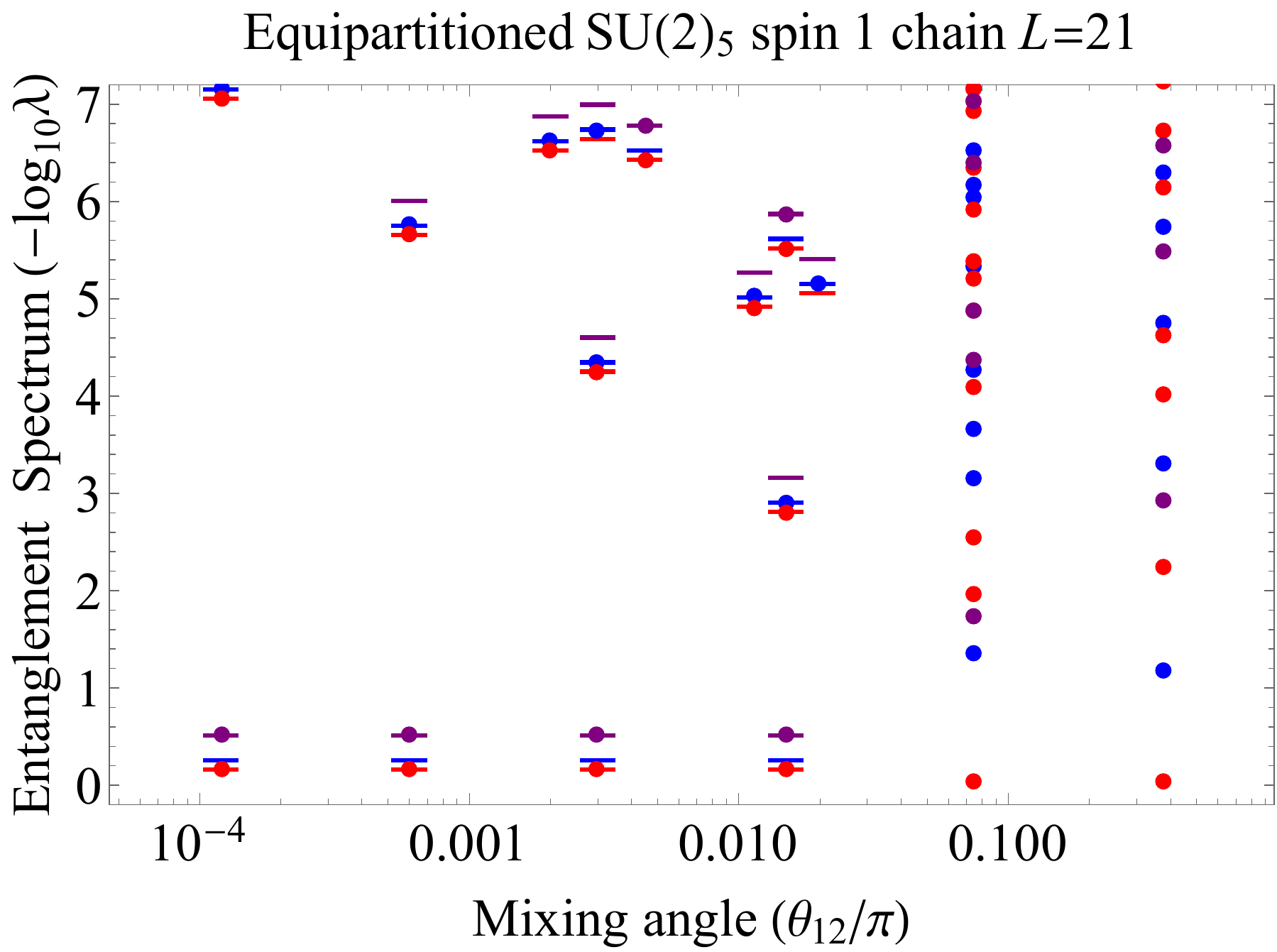}
	\caption{ES for the model Eq.~(\ref{Htheta}) as a function of $\theta_{12}$ parametrizing deviations from the AKLT sweet spot. Purple, red, and blue correspond to fusion channel $c=0,1,2$, respectively. Results are displayed in both linear (top) and logarithmic (bottom) scales. The multiplet structure of Eq.~(\ref{structure}) is evident within the topological phase, \textit{cf.} Fig~\ref{fig:1}.}\label{beyondSweetSpotversusTheta}
\end{figure}

\subsubsection{ES beyond the AKLT sweet spot} 
\label{se:awayfromsweeetspot}
\begin{figure}[t]
	\includegraphics[width=1\linewidth]{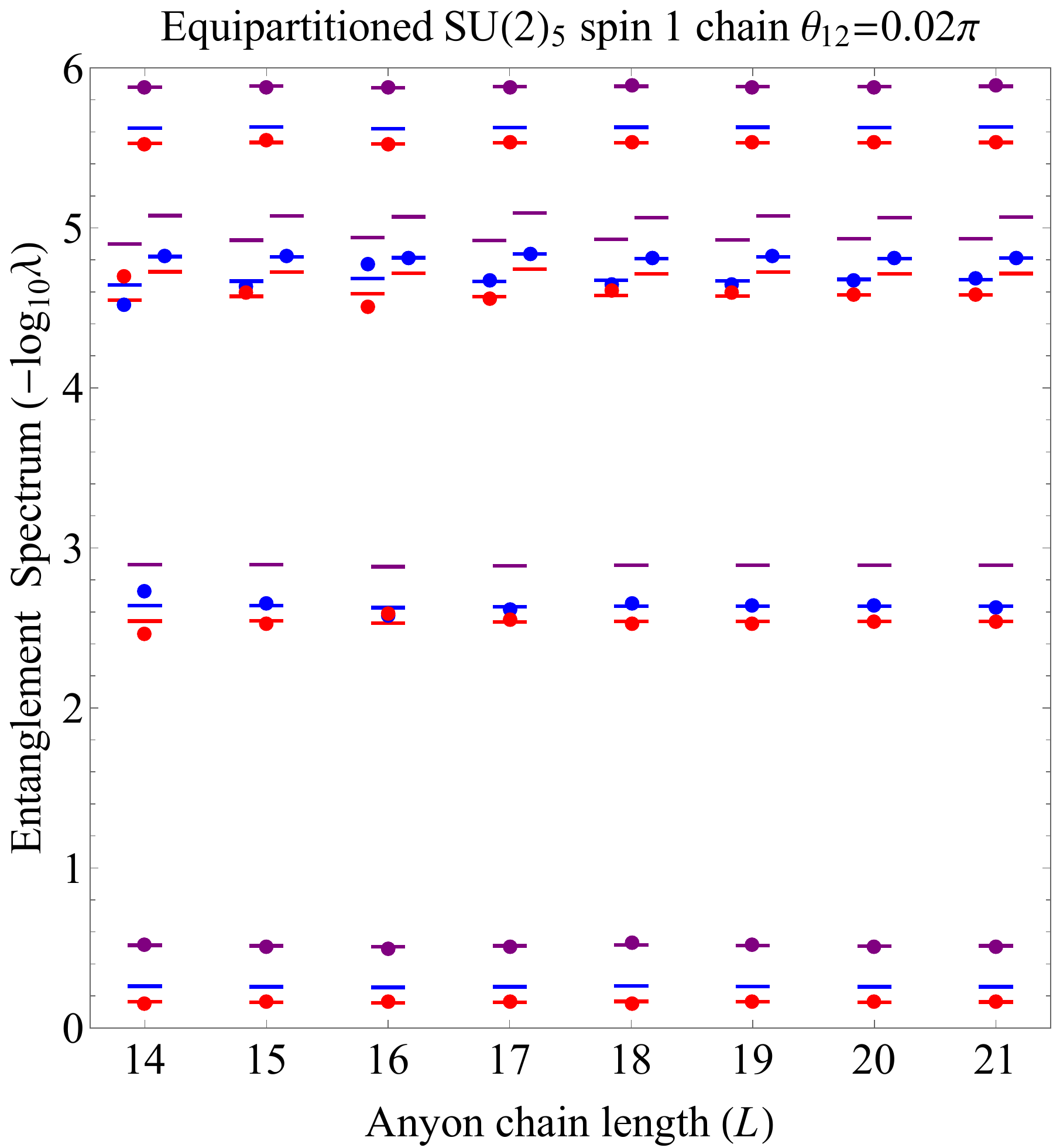}
	\caption{Convergence analysis of the results in Fig.~\ref{beyondSweetSpotversusTheta}.}\label{beyondSweetSpotversusL}
\end{figure}

Universal multiplet structure of the ES
described by the $SU(2)_k$ TQFT persists beyond the AKLT sweet spot, everywhere within the topological phase of the model  which extends approximately over the range of $-0.19\lesssim \theta_{12}/\pi \lesssim 0.06$~\cite{gils2013anyonic}. Our results for the ES, obtained  via exact diagonalization of the Hamiltonian followed by an SVD, are shown in Fig.~\ref{beyondSweetSpotversusTheta} as a function of $\theta_{12}$ for $L=21$. The large $L$ convergence of the various multiplets is demonstrated in Fig.~\ref{beyondSweetSpotversusL}. One can see that in addition to the dominant $c=0,1$ doublet at the AKLT sweet spot $\theta_{12}=0$, additional multiplets appear at finite $\theta_{12}$. This is similar to the behaviour of the ES in the Kitaev chain away from the sweet spot, \textit{cf.} Fig.~\ref{fig:1}(c).

The finite size results are in good agreement with the statement that the ratio between eigenvalues of states with charge $c$ within each multiplet is proportional to the quantum dimension $d_c$.

The results can be understood as a generalization of Eq.~(\ref{main}), which itself assumes well defined boundary anyons $a_L$, $a_R$. Starting from the OBC chain at finite $\theta_{12}$ at the ground state, and going in reverse through the steps in Fig.~\ref{fig:X}, one would obtain the diagram in Fig.~\ref{fig:X}(a) with two anyons lines connecting $A$ and $B$, \textit{cf.} Fig.~\ref{fig:Z}. Only at the AKLT sweet spot they carry a well defined charge $j=2$, however away from the AKLT sweet spot the state $| \Psi \rangle$ is a linear superposition over the possible values of $a_L$, $a_R$. Even in this case, we show in Appendix~\ref{app:general} that a multiplet structure emerges. Each multiplet is characterized by a pair of boundary anyons $a_L , \bar{a}_R$, with possible fusion outcome $c$ determined by their fusion rules. While the ground state multiplet corresponds to $(a_L , \bar{a}_R)=(2,2)$, the next multiplet and its multiplicity follows from $(a_L , \bar{a}_R)=(2,1)$ fusing into $c=1,2$ with unit multiplicity.

\section{Experimental tests} 
\label{se:exp}
Testing the anyonic-charge-resolved entanglement in an experiment requires (i) physical systems realizing non-Abelian anyons and (ii) protocols for practical measurement of quantum entanglement. In this section we discuss controllable experimental setups based on semiconductor quantum wires, which could realize Ising anyons. We propose to employ the schemes of Refs.~[\onlinecite{abanin2012measuring},\onlinecite{daley2012measuring}] to measure the $n-$th \Renyi{} entropy (RE), $s_n = {\rm{Tr}} \rho_A^n$, which were demonstrated in a cold-atom experiments~\cite{Islam2015Measuring,kaufman2016quantum}; for a recent work detailing this protocol, including the case of symmetry resoluiton, see Ref.~[\onlinecite{cornfeld2018imbalance}]. Our suggestion is that testing the universal internal structure of the ES can  be  achieved by extending these schemes to measure the charge-resolved entanglement~\cite{laflorencie2014spin,goldstein2017symmetry,PhysRevB.98.041106}.

The predicted universal structure in the ES of topological anyonic systems
is reflected in 
the  charge-resolved RE:
The separation of the ES into symmetry sectors $\{ \lambda\} = \cup_{c} \{ \lambda\}_{c}$ implies the additive structure of the RE, $s_n=  \sum_i \lambda_i^n  = \sum_{c} s_n(c)$. We refer to $s_n(c)$ as the charge-resolved RE.
Assuming that it 
is dominated by the largest eigenvalues in $\{ \lambda\}_{c}$, for which the low-energy TQFT result Eq.~(\ref{structure}) holds, we obtain the approximate relationship $s_n (c) \propto \#_{c} d_{c}^n$. Thus measuring the ratio between various charge-resolved contributions to the RE gives access to universal topological data.

In this section we use this simple observation to demonstrate, for systems realizing Ising anyons, that the known degeneracies in the ES can actually be measured. In this case $\#_{I} = \#_{\psi}=1$ and $d_I  = d_\psi$, yielding the familiar degeneracy in the ES, thus implying that the even and odd charged-resolved REs are equal. For Kitaev's Majorana chain model the degeneracy is symmetry protected thus this relation between the charged-resolved REs is exact. 

\begin{figure}[t] 
	\centering	
	\includegraphics[width=1\linewidth]{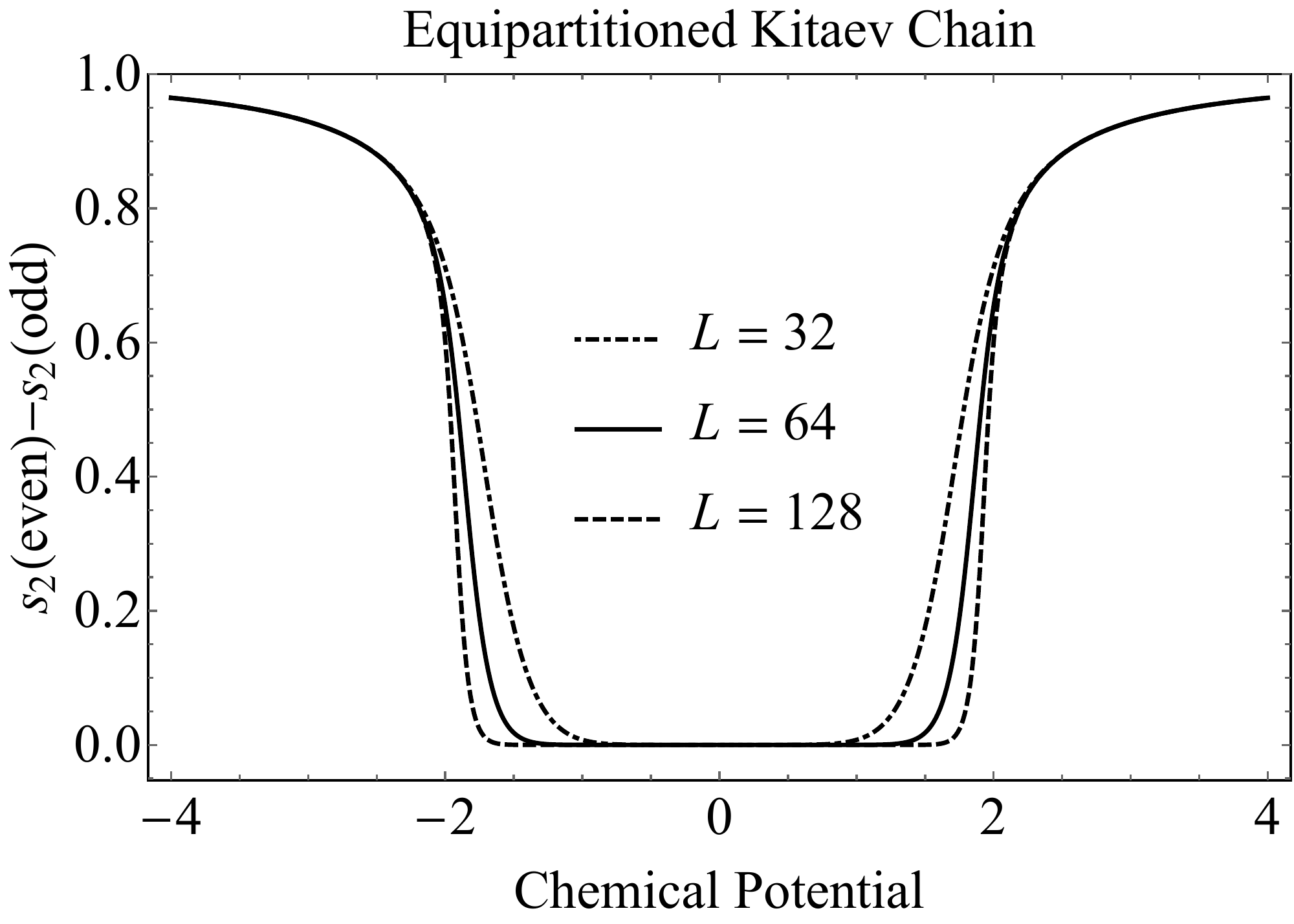}
	\caption{Parity resolved second \Renyi{} entropy for the Kitaev chain. Dependence of the difference $[s_{2}(I)-s_{2}(\psi)]$ on the chemical potential for an equipartition of an open chain with total length $L=32,64,128$, \textit{cf.} Fig.~\ref{fig:1}(c).
	} 
	\label{fig:s2es2o}
\end{figure}

This is explicitly shown in Fig.~\ref{fig:s2es2o}, where the same model parameters as in Fig.~\ref{fig:1}(c) are used to compute the parity resolved~\cite{goldstein2017symmetry,cornfeld2018imbalance} second REs, $s_2 (c=I) = s_2(\mathrm{even})$ and $s_2 (c=\psi) = s_2(\mathrm{odd})$, on which we concentrate in this section. One can see
that as the system enters the topological phase by tuning the chemical potential to the regime $|\mu| < 2$, the difference between the even- and odd-REs vanishes up to corrections exponentially small in the system size. 

Thus, the symmetry resolved measurement of the RE can serve as an order parameter for the topological phase: While the total entanglement entropy is just a non-universal number, when symmetry resolved, this number is exactly equipartitioned in the topological phase. 
		
We note, that a similar entanglement-based order parameter appears in the Schmidt gap~\cite{de2012entanglement} (\textit{i.e.}, the difference between the two largest eigenvalues of the reduced density matrix) as was found for the an equivalent transverse field Ising model, without explicitly decomposing the ES into symmetry sectors.

\subsection{Measurement of entanglement after fusion of Ising anyons in a quantum wire}
Consider a quantum wire
with spin-orbit coupling on top of which a floating superconductor is deposited. Upon application of a magnetic field a transition to topological superconductivity hosting Majorana end modes is expected~\cite{alicea2012new}, as suggested, \textit{e.g.}, in a recent experiment by Albrecht \textit{et~al}.~by the field evolution of the Coulomb blockade peaks~\cite{albrecht2016exponential}. This putative topological superconducting state supports Majorana fermion modes below the gap allowing either even or odd fermion parity. 

Aasen \textit{et~al}.~\cite{aasen2016milestones} envisioned interesting manipulations of this system allowing to entangle Majorana fermion modes. Consider the system in Fig.~\ref{fig:wire}(a) of a quantum wire Josephson-coupled to bulk superconducting leads. 
In the Coulomb blockade regime the even and odd ground states are split by Coulomb interactions, thus one may initialize an even ground state. Upon Josephson-coupling the wire to the bulk superconductors, charging effects become negligible, yielding an approximate even-odd degeneracy associated with the Majorana fermions~\cite{aasen2016milestones}. The switching-on of the Josephson coupling can be thought of as production of a pair of Majorana fermions denoted $\gamma_1, \gamma_4$ at the end of the wire, in the vacuum fusion channel. Now, a subsequent division of the wire into two segments denoted $A$ and $B$ by a central barrier, can be described as pulling the second pair of Majorana fermions $\gamma_2, \gamma_3$ from the vacuum. Identifying Majorana fermion modes with Ising anyons, the resulting anyonic wave function, depicted at the bottom of Fig.~\ref{fig:wire}(a), realizes the same state depicted in Fig.~\ref{fig:1}(a) with two pairs of Ising anyons produced from the vacuum. They create entanglement between two subsystems $A$ and $B$, which we wish to explicitly measure.

The resulting state can be written as $| 0_{14}0_{23}\rangle $, where $0_{ij}$($1_{ij}$) is the even (odd) fusion channel of Majorana fermion modes $\gamma_i$ and $\gamma_j$. This corresponds to the fusion of Ising anyons $\sigma \times \sigma$ with outcome $I$ ($\psi$). Expressing this state in the occupancy basis of two complex fermions $\gamma_1 +  i \gamma_2$ and $\gamma_3+i \gamma_4$ in regions $A$ and $B$, respectively, the same state
takes the form
\begin{equation}
\label{eq:Bell}
| 0_{14}0_{23}\rangle=\frac{1}{\sqrt{2}}\Big(| 0_{12}0_{34}\rangle +| 1_{12}1_{34}\rangle\Big).
\end{equation}
This realizes a maximally entangled Bell state between the left and right segments of the wire. 

To see the degeneracy of the ES explicitly, let us focus on the subsystem A - the left wire segment in Fig.~11(a), containing Majorana fermion modes  $\gamma_1$ and $\gamma_2$. Tracing out region B results in the reduced density matrix
\begin{equation}
\label{eq:bell}
\rho_A =\frac{1}{2}\Big(|  0_{12}\rangle \langle 0_{12} |+|  1_{12} \rangle \langle 1_{12} |\Big),
\end{equation}
implying $\lambda_I = \lambda_\psi=\frac{1}{2}$. 

In contrast to this entangled state, one may prepare a non-entangled state $\rho_A' = |  0_{12}\rangle \langle 0_{12} |$ by starting with two segments individually dominated by Coulomb interactions, hence having fixed charge, and only then opening the Josephson couplings to the bulk superconductors~\cite{aasen2016milestones}. 
Below we show how one can distinguish these states by explicitly measuring their parity-resolved second \Renyi{} entropy.

\subsection{Two-wire measurement setup}

\begin{figure}[t]
	\centering	
	\includegraphics[width=1\linewidth]{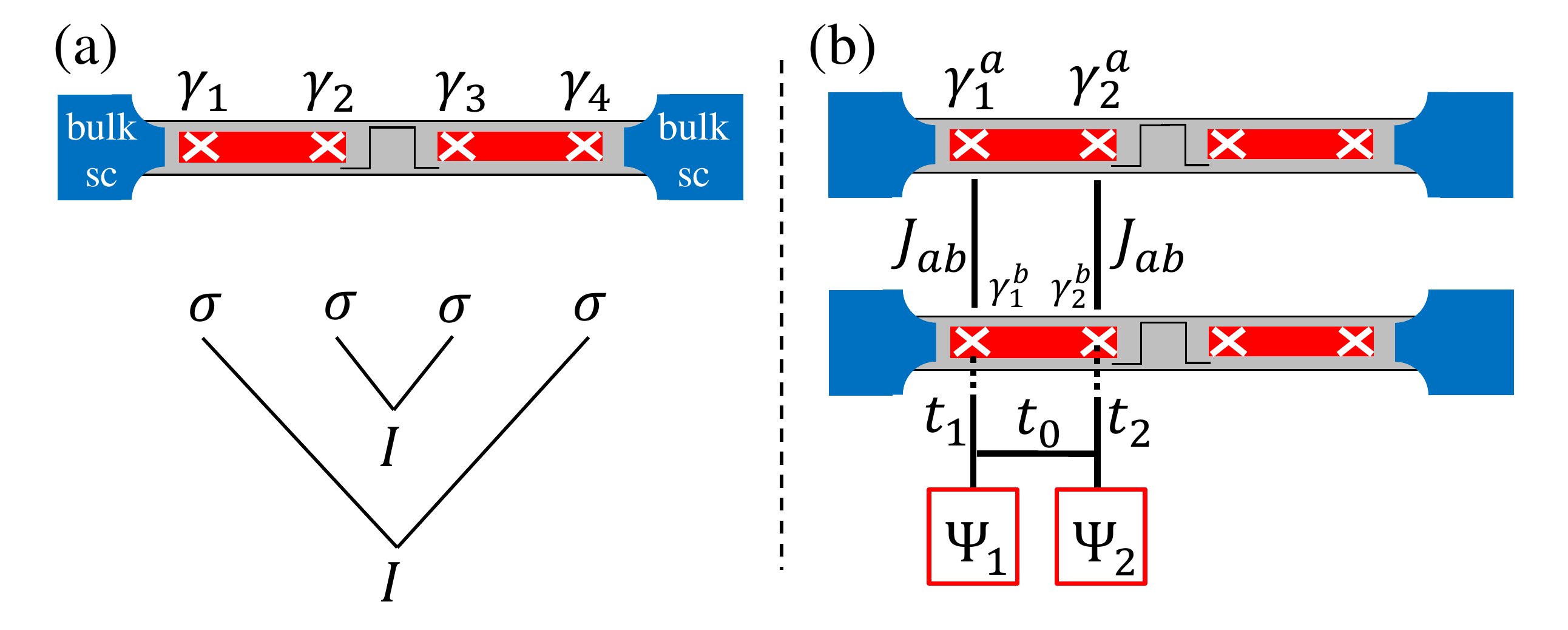}
	\caption{(a) Creation of an entangled state of Ising anyons in a quantum wire following Ref.~[\onlinecite{aasen2016milestones}]. Upon raising the central barrier, starting from a well define total charge state, one obtains the Bell state Eq.~(\ref{eq:Bell}) in which the parity of the left and right segments, comprising regions $A$ and $B$, respectively, are maximally entangled. (b) Two-copy system realizing a protocol to measure the parity-resolved second RE. The two copies are coupled by single particle tunnelling, $J_{ab}$, via the Majorana fermion modes, and then the parity of region $A$ is measured, \textit{e.g.}, by tunnel coupling ($t_1$) the Majorana fermion modes to two leads $\Psi_{1,2}$, as well as an interference arm $t_0$. A similar parity measurement device has to be connected to the top wire.
	} 
	\label{fig:wire}
\end{figure}

Following the protocol to measure the \Renyi{} entropy~\cite{daley2012measuring}, we add an identical wire hosting 4 Majorana modes, and prepare the same quantum
state in these wires. In the final state of the system, the Josephson couplings dominate over the charging energies so that charge may fluctuate. We use this to transfer charge between the wires via the Majorana modes. We label the two identical subsystem by $i = a, b$, and refer to them as copies. Subsystem $A$ in each copy has two Majorana operators $\gamma_1^{a/b}$ and $\gamma_2^{a/b}$. Thus for the Bell state Eq.~(\ref{eq:bell}) the 2-copy reduced density matrix is a product state
\begin{multline}
\rho_A^a \otimes \rho_A^b = \frac{1}{4}\Big(| 0_a 0_b \rangle \langle 0_a 0_b |+
| 0_a 1_b \rangle \langle 0_a 1_b |\\
+| 1_a 0_b \rangle \langle 1_a 0_b |+| 1_a 1_b \rangle \langle 1_a 1_b |\Big),
\end{multline}
where we dropped the Majorana indices for brevity. The next step of the measuring protocol is to
operate the following tunnelling Hamiltonian
\begin{equation}
H_{ab} = i J_{ab}(t)\gamma_1^a\gamma_1^b+i J_{ab}(t)\gamma_2^a\gamma_2^b,
\end{equation}
see Fig.~\ref{fig:X}(b). If we turn on this coupling for time $t = \frac{\pi \hbar}{4 J_{ab}}$ then the states of the system evolve into
\begin{eqnarray}\label{trans}
| 0_a 0_b \rangle &\mapsto& | 0_a 0_b \rangle,  \\
| 1_a 0_b \rangle &\mapsto& \frac{1}{\sqrt{2}}\Big(| 1_a 0_b \rangle + | 0_a 1_b \rangle \Big), \nonumber \\
| 0_a 1_b \rangle &\mapsto& \frac{1}{\sqrt{2}}\Big(| 1_a 0_b \rangle - | 0_a 1_b \rangle \Big), \nonumber \\ \nonumber
| 1_a 1_b \rangle &\mapsto& | 1_a 1_b \rangle.
\end{eqnarray}
Defining fermion $a$ as $\gamma_1^a = \frac{a^\dagger+a}{\sqrt{2}}$, $\gamma_2^a = \frac{a^\dagger-a}{\sqrt{2}i}$, and the same for copy $b$ with $a \leftrightarrow b$, this transformation is equivalent to~\cite{daley2012measuring}
\begin{equation}\label{fourier}
a \mapsto \frac{a+b}{\sqrt{2}},\qquad b \mapsto \frac{b-a}{\sqrt{2}}.
\end{equation}

\begin{table}[t]
    \begin{tabular}{cc||c|cc}

		$N_a$ &  $N_b$ & $f(N)$ & $f_{\mathrm{even}}(N)$ & $f_{\mathrm{odd}}(N)$ \\ \hline\hline
    0 & 0 & 1 & 1 & 0   \\ 
    0 & 1 & 0 & 0 & 0 \\ 
    1 & 0 & 0 & 0 & 0 \\ 
    1 & 1 & 1 & 0 & 1\\  
		
           \end{tabular}
            \caption{Measuring protocol for the second RE in  the even and odd parity fusion sector.\label{tab:1}}
        \end{table}

The final stage of the protocol is measuring the occupancy of subsystems $a,b$. Following Refs.~[\onlinecite{pichler2013thermal},\onlinecite{cornfeld2018measuring}] we find that the second RE, \(s_2=\Tr\rho_A^2\), is given by: (i) preparing the 2-wire setup, (ii) performing the (Fourier) transformation of Eqs.~(\ref{trans})~and~(\ref{fourier}), (iii) measuring the occupancies in each copy $N_a,N_b$,  and (iv) repeating steps (i-iii) and averaging over a function $f(N)$ where
\begin{equation}
f(N)=(-1)^{\frac{1}{2}(N_a+N_b)+N_a}\delta_{\frac{1}{2}(N_a+N_b)\in\mathbb{N}}=\delta_{\frac{1}{2}(N_a+N_b)\in\mathbb{N}},
\end{equation}
see Table~\ref{tab:1}.
The last equality is a consequence of having only two Majorana wires present in the system, hence the occupancies \(N_{a,b}\) are either $0$ or $1$.
Moreover, since the total occupancy is invariant under the transformation we may immediately read off the second RE for each fusion sector, \(s_2(\mathrm{even/odd})=\Tr[\rho_A^2\delta_{P,\mathrm{even/odd}}]\), where $(-1)^P\equiv(-1)^N$ is the parity operator ($P=0,1$). We thus find
\begin{equation}\label{majres}
f_{P}(N)=\delta_{\frac{1}{2}(N_a+N_b),P},
\end{equation}
see Table~\ref{tab:1}.

We now describe the occupancy measurement protocol for copy \(b\); copy \(a\) should be handled analogously.
Each one of the Majorana fermion modes
should be coupled to externally linked normal leads described by the Hamiltonian~\cite{landau2016towards,plugge2016roadmap} 
\begin{equation}
H_{\mathrm{parity}} = t_0 \Psi_1^\dagger \Psi_2+t_1\Psi_1^\dagger \gamma_1^b+t_2 \Psi_2^\dagger \gamma_2^b + \mathrm{H.c.},
\end{equation}
see Fig.~\ref{fig:wire}(b). This results in the effective tunnelling Hamiltonian
\begin{equation}
H_T=(t_0+i t_{12} \gamma_1^b \gamma_2^b) \Psi^\dagger_1 \Psi_2+\mathrm{H.c.},
\end{equation}
where $t_{12} \sim \frac{t_1 t_2}{E_c}$ with $E_c$ the charging energy. Measuring the conductance, it depends on the interference term $\sim t_0 t_{12} \langle  \gamma_1^b  \gamma_2^b \rangle $, allowing to measure the occupancy in $b$. Using Eq.~(\ref{majres}), one may average the measurement results $\{0,1\}$ to obtain the second \Renyi{} entropy.

\section{Summary and outlook} 
Using a topological quantum field theory approach, the entanglement spectrum of topological phases hosting non-Abelian anyons is found to have a universal structure: each eigenvalue can be labeled by the sub-system's anyonic charge, and its weight is dictated by the corresponding quantum dimension; likewise, its degeneracy is dictated by the fusion rules of the low energy topological theory. We tested our results for solvable anyon-chain models, specifically for the $SU(2)_k$ AKLT chains. Such gapped models elucidate the fact that as the subsystem size increases beyond the correlation length, the eigenvalues of the entanglement spectrum reach the predicted universal ratio dictated by quantum dimensions. Our result generalizes the well known degeneracies in the entanglement spectrum, \textit{e.g.},, in the Haldane chain, which serve as fingerprints of topological order, into universal ratios dictated by the quantum dimension of the possible fusion channels.

We have established a connection between our results on the entanglement spectrum, with a quantity that can be experimentally tested more easily. That is, measurements of the anyonic-charge-resolved \Renyi{} entropy provide information on the multiplet structure in the entanglement spectrum. We demonstrated this explicitly for Ising anyons realized in the Kitaev chain. Unfortunately this illustrates our results only in a partial way, since the quantum dimensions of the total topological charge correspond to Abelian anyons, hence the entanglement spectrum is characterized by degeneracies, and not by non-trivial ratios. Yet we have also shown that these degeneracies can be measured via the parity-resolved \Renyi{} entropy in Majorana wires. In principle it should be possible to generalize the symmetry resolved \Renyi{} entropy measurement protocol to arbitrary anyons by projecting anyon pairs onto specific fusion channels.

It would be interesting to connect our results with the ES of non-Abelian 2D FQH phases~\cite{Li2008}, which displays the structure of conformal towers describing 1D edge states (see Sec.~\ref{sec:intro} for further discussion and references). The derivation of our results assumes a simple wave function of few anyons fused from the vacuum (see Fig.~\ref{fig:1}); this is a natural description of 1D states, \textit{e.g.}, in the Kitaev chain or the various anyon-chain generalizations that we considered. However, the application of topological quantum field theory is not restricted to 1D. It also describes the non-Abelian sectors in 2D FQH phases, where the ES is quasi-continuous due to the gapless 1D edge states. Nevertheless, one can still label each state of the entanglement Hamiltonian by the total fusion channel of the subsystem. We expect the prediction that the ES contains universal multiplets dictated by the scaling dimensions and fusion rules to apply even in the presence of gapless edge states and leave this for future work.

Using our results, entanglement spectroscopy can be used to identify non-Abelian anyons as emergent particles in condensed matter systems of interacting bosons or fermions, both using computational methods, and with sufficient motivation, using experimental multi-copy entanglement-measurement methods.

\section*{Acknowledgements}
We thank Eddy Ardonne, Reinhold Egger, Masaki Oshikawa, Stephan Plugge, Simon Trebst, and Ari Turner for discussions and remarks. E.C.~acknowledges Chetan Nayak and the Boulder school for providing pedagogical background. K.S.~and E.S.~were supported by the US-Israel Binational Science Foundation (Grant No. 2016255).

\appendix
\section{Notations and normalizations of anyonic fusion algebras}\label{app:notation}
In this appendix we specify the notations and conventions of the anyonic fusion algebra in the paper.

The splitting space of two anyons, $a$ and $b$, with total charge, $c$, is a vector space $V_c^{ab}$ of dimension $N_{ab}^c$; it is indexed by $\mu=1,\dots,N_{ab}^c$. We label the states in this space either using bra-ket notation $\tket{a,b;c,\mu}\in V^{ab}_c$ or using the diagramatic presentation
\begin{equation}
\ket{a,b;c,\mu}=\left(\frac{d_c}{d_a d_b}\right)^{\frac{1}{4}}~\vcenter{\hbox{\includegraphics[height=7ex]{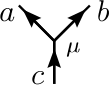}}},
\end{equation}
where the dual fusion space, $V_{ab}^{c}$, is spanned by
\begin{equation}
\bra{a,b;c,\mu}=\left(\frac{d_c}{d_a d_b}\right)^{\frac{1}{4}}\vcenter{\hbox{\includegraphics[height=7ex]{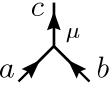}}}.
\end{equation}
We omit the $\mu$-index for $N_{ab}^c=1$ where it is redundant.
The benefits of the above diagram normalization is the orthonormality of the states:
\begin{gather}
\langle a',b';c',\mu'|a,b;c,\mu\rangle=\delta_{a,a'}\delta_{b,b'}\delta_{c,c'}\delta_{\mu,\mu'}\mathbb{I}_c,\\
\Tr\left\{\tket{a,b;c,\mu}\tbra{a',b';c',\mu'}\right\}=\delta_{a,a'}\delta_{b,b'}\delta_{c,c'}\delta_{\mu,\mu'}.
\end{gather}
The partial traces are similarly nicely normalized
\begin{equation}
\Tr_B\left\{\tket{a,b;c,\mu}\tbra{a',b';c',\mu'}\right\}=\delta_{a,a'}\delta_{b,b'}\delta_{c,c'}\delta_{\mu,\mu'}\mathbb{I}_a.
\end{equation}
For example, we may use these normalized states to resolve the identity operator
\begin{align}
\vcenter{\hbox{\includegraphics[height=10ex]{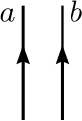}}}=\mathbb{I}_{ab}&=\sum_{c,\mu}\ket{a,b;c,\mu}\bra{a,b;c,\mu} \nonumber\\
&=\sum_{c,\mu}\sqrt{\frac{d_c}{d_a d_b}}~\vcenter{\hbox{\includegraphics[height=10ex]{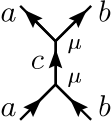}}}.
\end{align}

Note, that the relation between this ``bra-ket" normalization and Kitaev's $\psi$-notation~\cite{KITAEV20062} is $\tket{a,b;c,\mu}\Leftrightarrow(\tfrac{d_c}{d_a d_b})^{1/4}\psi^{ab}_{c,\mu}$. With this identification all diagrammatic representations are identical.
A complete description of anyonic fusion algebras in our normalized notation is found in Appendix A of Bonderson, Knapp, and Patel~\cite{BONDERSON2017399}; see also Ref.~[\onlinecite{bonderson2007non}].

\section{TQFT results for general edge content}\label{app:general}
The ES of anyonic models is complicated by two facts:
\subsection{Multiple boundary anyons}
First, one may have cases with multiple boundary anyons. In this case, the effective wave function is described in terms of anyons $\{a_i\}_{i=1}^n,\{\bar{a}_i\}_{i=1}^n$ fused from the vacuum 
and is given by
\begin{equation}
\rho= \prod_i\tket{a_i,\bar{a}_i;I}^L\tket{a_i,\bar{a}_i;I}^R\tbra{a_i,\bar{a}_i;I}^L\tbra{a_i,\bar{a}_i;I}^R.
\end{equation}
In this case, after tracing out region $A$, one obtains  (note that taking the partial trace is not equivalent to the partial anyonic trace~\cite{KITAEV20062,BONDERSON2017399}
\begin{equation}
\rho_A=\Tr_B{\rho}=\sum_{c}\frac{d_c}{1}[\Trt_B[\rho]_1]_c=\sum_{c} \frac{d_c}{\prod_i d_{a_i}^2} [\mathbb{I}_{\{a^L_i,\bar{a}^R_i\}}]_c,
\end{equation}
where in general the identity operator of the collection of anyons $\{a^L_i\}\cup\{\bar{a}^R_i\}$ to fuse into total channel $c$ contains $N_{\{a^L_i\}\cup\{\bar{a}^R_i\}}^c \ge 1$ many possibilities,
\begin{gather}
[\mathbb{I}_{\{a^L_i,\bar{a}^R_i\}}]_c=\!\!\!\!\sum_{\mu=1}^{N_{\{a^L_i,\bar{a}^R_i\}}^c}\!\!\!\tket{\{a^L_i,\bar{a}^R_i\};c,\mu}\tbra{\{a^L_i,\bar{a}^R_i\};c,\mu},\\
\mathbb{I}_{\{a^L_i,\bar{a}^R_i\}}=\sum_c [\mathbb{I}_{\{a^L_i,\bar{a}^R_i\}}]_c,\\ N_{\{a^L_i,\bar{a}^R_i\}}^c=\sum_{d^L,d^R}N_{\{a^L_i\}}^{d^L}N_{\{\bar{a}^R_i\}}^{d^R}N_{d^L d^R}^c,\\
N_{\{a_i\}}^{d}=\!\!\!\sum_{\{x_i\}_{i=2}^{n-1}}\!\!\! N_{a_1a_2}^{x_2}N_{x_2a_3}^{x_3}\cdots N_{x_{n-2}a_{n-1}}^{x_{n-1}}N_{x_{n-1}a_n}^{d}.
\end{gather}
This immediately yields the spectrum and degeneracies generalizing Eq.~\ref{main},
\begin{equation}
\label{main1}
\lambda_c=\frac{d_c}{\prod_i d_{a_i}^2},\qquad
\#_c=N_{\{a^L_i,\bar{a}^R_i\}}^c.
\end{equation}

\subsection{Superposition of anyons}
Second, the wave function may take the form of a linear superposition over the boundary anyons,
\begin{equation}
| \Psi \rangle = \sum_{a_L,\bar{a}_R} \alpha_{a_L,\bar{a}_R} \tket{a,\bar{a};I}^L \tket{a,\bar{a};I}^R.
\end{equation}
Consequently, the density matrix of the full system is
\begin{multline}
\rho=\sum_{a_L,\bar{a}_R} \alpha_{a_L,\bar{a}_R} \sum_{{a_L}',{\bar{a}_R}'} \alpha_{a_L',{\bar{a}_R}'}^\ast\\
\times\tket{a,\bar{a};I}^L\tket{a,\bar{a};I}^R\tbra{a',\bar{a}';I}^L\tbra{a',\bar{a}';I}^R.
\end{multline}
Tracing out subsystem $B$, as in Fig.~\ref{fig:secII}(c), brings a delta-function imposing $a_L=a_L'$, $a_R=a_R'$, and similarly for $\bar{a}_L$ and $\bar{a}_R$. Thus, 
\begin{equation}
\rho_A = \sum_{a_L,\bar{a}_R} |\alpha_{{a}_L,\bar{a}_R}|^2 \rho_A({a}_L,\bar{a}_R),
\end{equation}
where $\rho_A(a_L,\bar{a}_R)$ denotes the reduced density matrix for fixed boundary anyons.
In our main result, Eq.~(\ref{main}), various fusion trees associated with the possible internal fusion processes of the fixed boundary anyons lead to $\#_c$ distinct Schmidt states; see Eq.~(\ref{main1}). Thus, distinct values of the boundary anyons allow for a further distinction of the Schmidt states.
We conclude that each block of fixed $c$ admits a further block decomposition according to distinct values of $a_L,\bar{a}_R$, which together fuse into $c$,
\begin{equation}
\rho_A =\bigoplus_{{a}_L,\bar{a}_R} |\alpha_{{a}_L,\bar{a}_R}|^2  \bigoplus_{c\mathrm{~s.t.~}N_{a_L,\bar{a}_R}^c\neq 0}   \rho_A^{(c)}({a}_L,\bar{a}_R).
\end{equation}
We conclude that there are different multiplets labeled by two boundary anyons $a_L,\bar{a}_R$; Each multiplet decomposes according to the total anyon charge of region $A$, $c$, with a multiplicity given by $N_{\{a^L_i\}\cup\{\bar{a}^R_i\}}^c$ which indeed depends on the edge content. This allows to understand the results in Sec.~\ref{se:awayfromsweeetspot}.

\section{Exact ES at the AKLT sweet spot}\label{app:ESAKLT}
Here, we derive Eq.~(\ref{exactres}). The open BC anyon-chain wave-function takes the form
\begin{equation}
\psi =\sum_{\left\{x_i\right\}}f(\left\{x_i\right\})|x_1,\cdots,x_L\rangle,
\end{equation} 
where~\cite{gils2013anyonic} $f(\left\{x_i\right\})=(-1)^{\#1}(d_1)^{(\#1-L)/2}(d_2)^{L/2-\#1}$, $L$ is the number of {\it inner} anyonic legs, $\#j$ is the number of appearance of $j-$type anyons, and we impose $x_i\in\{0,1\}$ and $x_1=x_L=0$. This wave function is not normalized; this is later taken into account. The form of this wave function dictates that $\lambda_2=0$ for all $L$, since none of the anyons can be a $2-$type anyon.

Let us define $\chi_{ij}(n)=\sum_{\left\{x_i\right\}\neq x_1,x_n}f^2(\left\{x_i\right\})_{x_1=i,x_{n}=j}$ to be the probability to find a string with $n$ inner legs in which the first and $n\mathrm{th}$ anyons are $x_1=i$, $x_n=j$. We also define $\chi_{i}(1)=f^2(i)$ to be squared coefficient of a single leg anyon string of length $1$. The system under consideration is a bi-partitioned string with $n_1$, $n_2$ {\it outer} anyons in each sector, see Fig.~(\ref{fig:fusiondiagrammodel}). Since the singular values are norm preserving, we can represent the two unnormalized values of the entanglement spectrum as a product of two strings
\begin{eqnarray}
(\bar{\Lambda}_0)^2&=&\frac{\chi_{00}(n_1+1)\chi_{00}(n_2+1)}{\chi_{0}(1)},  \\  \nonumber
(\bar{\Lambda}_1)^2&=&\frac{\chi_{01}(n_1+1)\chi_{10}(n_2+1)}{\chi_{1}(1)},
\end{eqnarray}
where the factors $\chi_{0}(1)$  ($\chi_{1}(1)$) in the denominator are due to double counting of a single $0$ ($1$) anyon. The bar sign indicates that these entanglement spectrum values are not normailzed. However, we can take care of normalization by observing that the string $\chi_{00}(n_1+n_2+1)$ includes both $\bar{\Lambda}_0$ and $\bar{\Lambda}_1$. Therefore, the normalized values are
\begin{eqnarray}
(\Lambda_0)^2&=&\frac{\chi_{00}(n_1+1)\chi_{00}(n_2+1)}{\chi_{0}(1)\chi_{00}(n_1+n_2+1)},  \\  \nonumber
(\Lambda_1)^2&=&\frac{\chi_{01}(n_1+1)\chi_{10}(n_2+1)}{\chi_{1}(1)\chi_{00}(n_1+n_2+1)}. 
\end{eqnarray}
Using the string identities $\chi_{01}(n)=\chi_{0}(1)\chi_{11}(n-1)$ and $\chi_{00}(n)=\chi_{00}(2)\chi_{11}(n-2)$ we rewrite the above equations as
\begin{eqnarray}
\label{chis}
(\Lambda_0)^2&=&\frac{\chi_{0}(1)\chi_{11}(n_1-1)\chi_{11}(n_2-1)}{\chi_{11}(n_1+n_2-1)}, \\ \nonumber
(\Lambda_1)^2&=&\frac{\chi_{11}(n_1)\chi_{11}(n_2)}{\chi_{1}(1)\chi_{11}(n_1+n_2-1)}.
\end{eqnarray}

Consequently, in order to obtain a closed form of the entanglement spectrum for all $n$ we need to explicitly calculate only the value of $\chi_{11}(n)$. To do so, we observe that each $\chi_{11}(n)$, which is nothing but an open string starting and ending with $1$ anyons, can be recursively expressed as  
\begin{eqnarray}
\chi_{11}(n)&=&\chi_{1}(1)\chi_{11}(n-1)+\chi_{10}(2)\chi_{11}(n-2),  \nonumber \\
&=&\frac{1}{d_1}\chi_{11}(n-1)+\frac{1}{d_2}\chi_{11}(n-2). 
\end{eqnarray}
This equation has a simple closed form solution:
\begin{eqnarray}
\chi_{11}(n) =A+\frac{B}{(-d_1)^n},
\end{eqnarray} 
where $A$ and $B$ are some constants. Using  $\chi_{11}(2)=\frac{1}{d_2^2}$ and the value of $\chi_{1}(1)=\frac{1}{d_2}$, we find 
\begin{eqnarray}
\chi_{11}(n)&=&\frac{d_{1}}{d_{2}^{3}}\left[1-\frac{1}{\left(-d_{1}\right)^{n}}\right].  
\end{eqnarray} 
Plugging this at Eq.~(\ref{chis}) we obtain the general form of the entanglement spectrum
\begin{eqnarray}
\Lambda_{0}&=&\frac{1}{d_{2}}\sqrt{\frac{\left[1-\left(-d_{1}\right)^{-n_1+1}\right]\left[1-\left(-d_{1}\right)^{-n_2+1}\right]}{\left[1-\left(-d_{1}\right)^{-n_1-n_2+1}\right]}}, \nonumber \\ 
\Lambda_{1}&=&\frac{\sqrt{d_{1}}}{d_{2}}\sqrt{\frac{\left[1-\left(-d_{1}\right)^{-n_1}\right]\left[1-\left(-d_{1}\right)^{-n_2}\right]}{\left[1-\left(-d_{1}\right)^{-n_1-n_2+1}\right]}}.
\end{eqnarray}\\
For equally partitioned system $n_1=n_2=n$ one obtains
\begin{eqnarray}
\Lambda_{0}&=&\frac{1}{d_{2}}\cdot\frac{1-\left(-d_{1}\right)^{-n+1}}{\sqrt{1+d_{1}^{-2n+1}}},  \\ \nonumber
\Lambda_{1}&=&\frac{\sqrt{d_1}}{d_{2}}\cdot\frac{1-\left(-d_{1}\right)^{-n}}{\sqrt{1+d_{1}^{-2n+1}}}.
\end{eqnarray}
This completes the derivation of the ES $\{ \lambda \}=\{ \Lambda^2 \}$ in Eq.~(\ref{exactres}).

\bibliographystyle{apsrev4-1}

\end{document}